\documentclass[10pt]{iopart}

\usepackage{iopams}
\usepackage{dsfont}
\usepackage{graphicx}
\usepackage{dcolumn}
\usepackage{bm}
\usepackage{epsfig}

\newcommand{\be}{\begin{eqnarray}}
\newcommand{\ee}{\end{eqnarray}}
\def\refeq#1{(\ref{#1})}
\def\vec#1{\mathbf#1}
\def\d{\mbox d}
\def\wt{\widetilde}
\def\nn{\nonumber}
\def\i{\int_{-\infty}^{\infty}}
\def\ip{\int_{0}^{\infty}}

\def\G{\Gamma}

\def\la{\lambda}

\def\g{\gamma}
\def\al{\alpha}
\def\s{\sigma}
\def\e{\epsilon}
\def\o{\omega}

\def\l{\left}
\def\r{\right}
\def\te{\mbox{e}}
\def\rmi{{\rm{i}}}

\def\up{\uparrow}
\def\down{\downarrow}
\def\ovj{{\overline J}}

\def\H{{\mathcal H}}

\begin{document}

\title[The one-dimensional Hubbard model with open ends]{The one-dimensional
  Hubbard model with open ends: Universal divergent contributions to the
  magnetic susceptibility} 
\author{Michael Bortz} 
\address{Department of
  Theoretical Physics, Research School of Physics and Engineering, Australian
  National University, Canberra ACT 0200, Australia} \author{Jesko Sirker}
\address{Department of Physics and Astronomy, University of British Columbia,
  Vancouver, B.C., Canada V6T 1Z1}


\begin{abstract}
  The magnetic susceptibility of the one-dimensional Hubbard model with open
  boundary conditions at arbitrary filling is obtained from field theory at
  low temperatures and small magnetic fields, including leading and
  next-leading orders. Logarithmic contributions to the bulk part are
  identified as well as algebraic-logarithmic divergences in the boundary
  contribution. As a manifestation of spin-charge separation, the result for
  the boundary part at low energies turns out to be independent of filling and
  interaction strength and identical to the result for the Heisenberg model.
  For the bulk part at zero temperature, the scale in the logarithms is
  determined exactly from the Bethe ansatz. At finite temperature, the
  susceptibility profile as well as the Friedel oscillations in the
  magnetisation are obtained numerically from the density-matrix
  renormalisation group applied to transfer matrices. Agreement is found with
  an exact asymptotic expansion of the relevant correlation function.
\end{abstract}

\section{Introduction}
The one-dimensional Hubbard model plays a central role in the understanding of
interacting electrons in one dimension. The Hamiltonian 
\be 
\fl H=-\sum_{j=1}^{L-1}
\sum_{a=\up,\down}\l(c^\dagger_{j,a} c_{j+1,a} + c^\dagger_{j+1,a}
c_{j,a}\r)+4 u \sum_{j=1}^Ln_{j,\up}n_{j,\down}-\frac{h}{2} \sum_{j=1}^L\l(n_{j,\up}-n_{j,\down}\r)\label{hdef}.  
\ee 
arises
naturally in the tight-binding approximation of electrons on a chain with $L$
sites. In \refeq{hdef}, the magnetic field $h$ couples to the $z$-component of the total spin. The interaction parameter $u=U/|t|>0$ is the ratio between the on-site
Coulomb repulsion $U$ and the hopping amplitude $t$. Note that the eigenvalues of $H$ are invariant under a sign change $t\to -t$ \cite{lieb68}. Furthermore, $H$ is invariant under reversal of all spins and under a particle-hole transformation (the so-called Shiba transformation) \cite{lieb68}. Therefore we restrict ourselves here to positive magnetisation and lattice filling less than or equal to one. 

The appealing simplicity
of the Hamiltonian, combined with its Bethe-Ansatz solvability and its
adequateness for field-theoretical studies are the reasons for its wide
popularity.

Recent experimental achievements in two areas additionally motivate our studies: On the one hand, the fabrication and characterisation of carbon
nanotubes have shown that these can be considered as realisations of one-dimensional Hubbard models \cite{mazz00}.
Especially, kinks in these quantum wires have been realised experimentally
\cite{yao99,iij96}. A kink locally weakens the hopping amplitude at one specific
lattice site in the Hamiltonian. Such a modification is known to
be a relevant perturbation, which, at $u>0$, is governed by a fixed point with zero
conductance through the kink \cite{KaneFisherPRB}. 
Thus at low energies the chain is effectively cut into two pieces.

On the other hand, ultracold fermionic atoms in optical lattices can be
described by a one-band Hubbard model \cite{hof02}. Given the recent progress
in realizing quasi one-dimensional bosonic quantum gases \cite{par04}, it is
likely that similar experimental progress will be made with fermions.

In order to model these situations, we consider open boundary conditions in
\refeq{hdef}. Compared to the case with periodic boundary conditions, an
additional surface contribution $f_B$ to the free energy occurs, defined as
\be f_B&=& \lim_{L\to \infty} \l(F_{obc}-F_{pbc}\r), \ee where $F_{obc}$
($F_{pbc}$) is the total free energy for open (periodic) boundary conditions. In this work, we will focus on the magnetic susceptibility per lattice site 
$\chi=\chi_{{\rm{bulk}}}+ \chi_B/L$ both at $T=0$, $h\geq 0$ and $T\geq 0$,
$h=0$, at arbitrary fillings. 

For half filling, the bulk contribution $\chi_{\rm{bulk}}$ at $T=h=0$ has first been given by Takahashi \cite{tak69}, where also the existence of logarithmic contributions at finite $h$ is mentioned. Later on, Shiba \cite{shi72} calculated $\chi_{\rm{bulk}}$ at $T=h=0$ for general filling. The free energy at finite temperatures was given by Takahashi (for an overview and original literature, cf. the book \cite{tak99}), and later by Kl\"umper (the book \cite{book} contains a detailed account of this work). However, it seems as if the explicit behaviour of $\chi_{\rm{bulk}}$ at $T=0$, $h\gtrsim 0$ and $T\gtrsim 0$, $h=0$ has not been derived so far. In this work, this gap will be filled.   

Although the boundary quantity $\chi_B$ is only an $\mathcal O(1)$-correction
to the total bulk contribution, it may become important in experiments if it
shows a divergency with respect to the temperature $T$ or magnetic field $h$.
Indeed, such divergences have been discovered and analysed in the spin-1/2
Heisenberg chain
\cite{FujimotoEggert,FurusakiHikihara,AsakawaSuzuki96a,BortzSirker,sir05b}.
Since the isotropic spin-1/2 Heisenberg chain is obtained from \refeq{hdef} in
the limit $u\to \infty$, related divergences are also expected to occur in
\refeq{hdef}. In the case of half-filling for $T=0$, it has in fact been shown
in \cite{AsakawaSuzuki96a,AsakawaSuzuki96b} that the boundary magnetic
susceptibility is divergent, $\chi_B\sim 1/(h\ln^2 h)$, for $h\to 0$.  Exactly
the same result for $\chi_B$ has also been obtained for the supersymmetric
$t-J$ model \cite{FHLE}. The OBCs do not only lead to $1/L$-corrections but
also break translational invariance. Therefore quantities like the
magnetisation or the density become position dependent.  Local measurements of
such quantities then provide a way to obtain information about the impurity
making theoretical predictions about the behaviour of such one-point
correlation functions desirable.

In section \ref{ft} we give the functional forms of both the bulk and the
boundary contributions by using a field-theoretical argument. Leading and
next-leading logarithmic contributions to the finite bulk susceptibility are
found both at finite $T$ and finite $h$. On the other hand, the boundary
contribution diverges in a Curie-like way with logarithmic terms, where again
we give both the leading and next-leading divergences. These results depend
each on two constants, which are the spin velocity and the scale involved in
the logarithms. At magnetic fields or temperatures much smaller than this
scale the result for the boundary susceptibility becomes independent of the
spin velocity and the scale and therefore completely universal. This is a
manifestation of spin-charge separation as will become clear in the next
section. For the bulk susceptibility, on the other hand, only the functional
dependence on field or temperature will be universal for low energies. The
value at zero field and zero temperature, however, is a non-universal constant
which does depend on filling and interaction strength via the spin velocity.

The spin velocity has been determined previously from the Bethe ansatz
solution \cite{shi72}. In section \ref{ba}, for $T=0$, the scale in the
logarithms will be determined exactly as well. The calculation of boundary
effects at $T>0$ based on the Bethe ansatz solution still remains an open
problem, as for all Yang-Baxter integrable models (reasons for that are given
in \cite{BortzSirker,GoehmannBortz} for the special case of the spin-1/2 $XXZ$
chain). In section \ref{corrfkts} we therefore calculate the susceptibility
profile $\chi(x,T)$ and magnetisation profile $s(x,T)$ in the asymptotic
low-energy limit (that is, for large distances and small temperatures) by
making use of conformal invariance. Due to the open boundaries, $s(x,T)$ shows
the characteristic Friedel oscillations \cite{egg95,fab95,bed98}. To test the
field-theory predictions, we perform numerical calculations in the framework
of the density matrix renormalisation group applied to transfer matrices,
which is particularly suited for impurity and open-boundary models. In the
last section we will present our conclusions and discuss in which experimental
situations the calculated boundary effects might become important.

\section{The low-energy effective Hamiltonian}
\label{ft}
First, we briefly review the effective field theory for the Hubbard model
following in large parts Refs.~\cite{affreview,book}. From the effective
Hamiltonian we then obtain the magnetic susceptibility at small magnetic field
and low temperature.

Let $a$ be the lattice spacing. We introduce fermionic fields $\psi(x)$ 
in the continuum by
\be
c_{j\sigma}\to \sqrt{a}\psi(x) = \sqrt{a}(\te^{\rmi k_{F\sigma}x} R_\sigma(x)
+ \te^{-\rmi k_{F\sigma}x} L_\sigma(x))
\label{split}
\ee where $x=j\cdot a$ and the usual splitting into left- and right-moving
parts has been performed. The Fermi momentum depends on both the density $n$
and the magnetisation $s$ as $k_{F\up}=\pi(n+2s)/(2 a)$,
$k_{F\down}=\pi(n-2s)/(2 a)$ (the magnetisation is defined as
$s=(m_\up-m_\down)/2$, with $m_\sigma=M_\sigma/L$ being the density of
particles with spin $\sigma$). In the following, we will consider the
zero-field case where $k_{F\up}=k_{F\down}=\pi n/(2 a)$. Eq.~\refeq{split}
allows it to introduce a Hamiltonian density $\H(x)$, such that $H= \int \H \d
x$. In terms of the right- and left-movers in (\ref{split}) the kinetic part of
the Hamiltonian \refeq{hdef} in a lowest order expansion in $a$ becomes 
\be
\label{eff1}
\H_0 =-\rmi v_F \sum_\sigma \l( R_\sigma^\dagger\partial_x R_\sigma -
L_\sigma^\dagger\partial_x L_\sigma \r)
\ee
and the interaction part
\be
\label{eff2}
\fl\H_{int}= 4ua\l\{ :\l(R_\uparrow^\dagger R_\uparrow + L_\uparrow^\dagger
L_\uparrow\r)\l(R_\downarrow^\dagger R_\downarrow + L_\downarrow^\dagger
L_\downarrow\r) :\r. - : R_\uparrow^\dagger R_\downarrow L_\downarrow^\dagger L_\uparrow : - :
R_\downarrow^\dagger R_\uparrow L_\uparrow^\dagger
L_\downarrow : \nn \\
\fl\qquad\qquad- \l.\l(\te^{4\rmi k_Fx} L_\uparrow^\dagger L_\downarrow^\dagger R_\uparrow
R_\downarrow + h.c.\r)\r\} \,.\ee 
Here ``:'' denotes normal ordering. For brevity, the $x$-dependence of the
operators has been dropped. The Fermi velocity is given by $v_F:= 2 a \sin(k_F
a)$. The second term in \refeq{eff2} represents backward scattering processes
whereas the last term is due to Umklapp scattering. Only in the half-filled
case, where $k_F=\pi/(2a)$, is the Umklapp term non-oscillating and has to be kept
in the low-energy effective theory. For all other fillings it can be dropped.

To make the
Hamiltonian manifestly $SU(2)$ invariant under spin-rotations one can also
express $\H$ in terms of the following currents
\cite{Affleck_SU(N),book,tsvbook}: 
\be
J=\sum_\sigma :R^\dagger_\sigma R_\sigma: \qquad ,\qquad \ovj = \sum_\sigma :L^\dagger_\sigma L_\sigma:\label{jdef}\,,\nn\\
J^a = \frac12 \sum_{\alpha,\beta}: R^\dagger_\alpha \sigma_{\al\beta}^a
R_\beta: \qquad , \qquad \ovj^a = \frac12 \sum_{\alpha,\beta}:
L^\dagger_\alpha \sigma_{\al\beta}^a L_\beta: \,.\nn
\ee 
The free part \refeq{eff1} then reads
\be \H_0&=& v_F\l[
\frac\pi2 \l( :J^2: + :\ovj^2: \r) + \frac{2\pi}{3} \l( :\vec J \cdot \vec J :
+ :\vec \ovj \cdot \vec \ovj:\r) \r]\label{h0}. \ee 

As far as the interaction part \refeq{eff2} is concerned, we first consider
the case $n\neq 1$, that is away from half-filling. Then Umklapp scattering
can be ignored leading to 
\be
\fl\H_{int}=  g_c\l[ :J^2: + :\ovj^2:\r] + g_s \l[ :\vec J \cdot \vec J : + :\vec \ovj \cdot \vec \ovj:\r] + \lambda_c :J \ovj: + \lambda :\vec J \cdot \vec \ovj:\label{hint}
\ee
and coupling constants $g_c = ua $, $g_s = -4 ua /3$, $\lambda_c=2 ua $ and
$\lambda=-8 ua $. 

Taking Eqs.~\refeq{h0}, \refeq{hint} together, we see that the Hamiltonian is a
sum of two terms: one depending on the scalar currents $J,\,\ovj$ only
(corresponding to charge excitations) and the second depending on the vector
currents $\vec J$, $\vec \ovj$ (associated with spin excitations). 
\be
\mathcal H_c& =& \l(\frac{\pi v_F}{2}+g_c\r) :\l[ J^2  + \ovj^2\r]:  + \lambda_c : J \,\ovj:  .\label{hc0}\\
\mathcal H_s& =& \l(\frac{2\pi v_F}{3}+g_s\r)  :\l[ \vec J \cdot \vec J + \vec \ovj \cdot \vec \ovj\,\r]:  + \lambda  : \vec J \cdot \vec \ovj:  .\label{hs}
\ee
The charge and spin parts commute, $\l[\mathcal H_c,\mathcal H_s\r]=0$. 

Let us first focus on $\mathcal H_c$. The term with coefficient $g_c$ gives rise to a
renormalisation of $v_F$, yielding the charge velocity 
\be v_c=v_F+2 ua/\pi
\label{vc}.  
\ee 
Upon bosonising, the charge currents are written as
\cite{affreview} 
\be J&=& -\frac{1}{\sqrt{4\pi}}\l(\Pi + \partial_x
\varphi\r),\;\ovj = \frac{1}{\sqrt{4\pi}}\l(\Pi - \partial_x \varphi\r)\nn 
\ee
where the boson field $\varphi$ and the corresponding momentum $\Pi$ satisfy
the canonical commutation relation $\l[\varphi(x),\Pi(x')\r]=\rmi
\delta\l(x-x'\r)$. Then 
\be \mathcal H_c&=& \frac{v_c}{4}\l[
\Pi^2\l(1-\frac{\lambda_c}{\pi v_c}\r) + \l(\partial_x \varphi\r)^2
\l(1+\frac{\lambda_c}{\pi v_c}\r)\r]\label{hc}.  
\ee 
By scaling
$\varphi'=\varphi \sqrt{K_c}$, $\Pi'=\Pi/\sqrt{K_c}$, 
\be
K_c=1-\frac{\lambda_c}{\pi v_c}\approx 1-\frac{2ua}{\pi v_F}\label{kc}, 
\ee 
this
Hamiltonian is written as $\mathcal H_c= \frac{v_c}{4}\l[ \Pi'^2 +
\l(\partial_x \varphi'\r)^2 \r]$, which describes noninteracting fields. Note,
that in this field-theoretical approach the Luttinger parameter $K_c$ is calculated only up to the linear order in $u$. The same is true for
$v_{c,s}$. Contributions in higher
$u$-order would occur if the perturbational approach is pursued further.
Fortunately, the Bethe-ansatz solvability of the model allows it to calculate
$v_{c,s}, K_c$ exactly \cite{book}. We will come back to this point in the
next section.

In $\mathcal H_s$, the $g_s$-term leads to a renormalisation of the spin velocity 
\be
v_s=v_F-2 ua/\pi\label{vs}.
\ee
The remaining interaction of vector currents is a marginal perturbation. By setting up the corresponding renormalisation group equations, it turns out that it is marginally irrelevant (relevant) for sgn$(\lambda)<0$ (sgn$(\lambda)>0$) \cite{affreview,zam95}. In our case, $\lambda=-8 ua < 0$. The remarkable point about this is that the spin part of the low-energy effective Hubbard model is identical to the corresponding expression for the $XXX$-Heisenberg chain \cite{egg94,luk98}, whereas the charge part is described by free bosons (away from half filling). For this case, field theory has been employed \cite{egg94,luk98,zam95} to obtain the bulk contribution to the magnetic susceptibility in the form
\be
\fl\chi_{{\rm{bulk}}}(E)=\chi_0\l(1+\frac{1}{2 \ln E_0/E} - \frac{\ln\ln E_0/E}{4 \ln^2 E_0/E}+ \frac{\g_E}{\ln^2 E_0/E}+\ldots\r)\label{suszi}\\
\fl\chi_0= \frac{1}{2\pi v_s}\label{chi0}, 
\ee
where $E=h,T$ (magnetic field or temperature), $E_0=h_0,T_0$ is some scale
and $\chi_0:= \chi(T=0,h=0)$ is given by the inverse of the spin velocity. For
the open $XXX$-chain, the boundary contributions have been found to be
\cite{FujimotoEggert,FurusakiHikihara,sir05b}: 
\be
\chi_{B}(T)&=&\frac{1}{12 T \ln T_0^{(B)}/T}\l(1-\frac{\ln \ln T_0^{(B)}/T}{2 \ln T_0^{(B)}/T}+\frac{\g_T^{(B)}}{\ln T_0^{(B)}/T}+ \ldots\r)\label{boun1}\\
\chi_{B}(h)&=&\frac{1}{4 h \ln^2 h_0^{(B)}/h}\l(1-\frac{\ln \ln h_0^{(B)}/h}{\ln(h_0^{(B)}/h)}+\frac{\g_h^{(B)}}{\ln h_0^{(B)}/h}+\ldots\r)\label{boun2}
\ee
From the above considerations we conclude that the bulk and boundary
contributions to the magnetic susceptibility in the Hubbard model are also of
the form (\ref{suszi})-(\ref{boun2}), where, compared to the
$XXX$-model, $\chi_0$, $T_0$, $h_0$, $\g_{T,h}$, $\g^{(B)}_{T,h}$ are
renormalised by the charge part. 

Most interestingly, the pre-factor of $\chi_B$ remains unaffected by
the charge channel. The divergent boundary contribution for $T\ll T_0^{(B)}$ or
$h\ll h_0^{(B)}$ is therefore completely universal
\be
\chi_{B}\l(T\ll T_0^{(B)}\r)&=&-\frac{1}{12 T \ln T}\l(1+\frac{\ln|\ln
  T|}{2\ln T}\r)\label{uni1}\\
\chi_{B}\l(h\ll h_0^{(B)}\r)&=&\frac{1}{4 h \ln^2 h}\l(1+\frac{\ln|\ln h|}{\ln
  h}\r)\label{uni2} \; .
\ee
This can be understood as follows: As the charge- and
spin-part of the Hamiltonian separate at low energies, the only way how the
charge channel can effect the spin channel is by a renormalisation of $v_s$
and $K_s$. The Luttinger parameter $K_s$ in the spin sector, however, is fixed
to $K_s\equiv 1$ due to the $SU(2)$ symmetry and cannot renormalise. The
explicit calculations of the pre-factor of $\chi_B$ for the $XXX$-model in
\cite{FujimotoEggert,FurusakiHikihara,sir05b} show, on the other hand, that it does
not depend on the spin-velocity $v_s$. It therefore remains completely independent
of the filling factor and the interaction strength.

Let us now comment on the scales involved in Eqs. \refeq{suszi}, \refeq{boun1}, \refeq{boun2}. Including the order $\mathcal O(\ln^{-2} E)$ in Eq.~\refeq{suszi}, this equation can be written as
\be
\chi_{{\rm{bulk}}}(E)&=&\chi_0\l(1-\frac{1}{2 \ln E} - \frac{\ln|\ln E|}{4 \ln^2 E}+ \frac{\g_E-(\ln E_0)/2}{\ln^2 E}+\ldots\r).
\ee
It is convenient to define a new scale $E_0= \wt E_0 \te^{2 \g_E}$. Then, again up to the order $\mathcal O\l(\ln^{-2}E\r)$, we have 
\be
\chi_{{\rm{bulk}}}(E)&=&\chi_0\l(1+\frac{1}{2 \ln \wt E_0/E} - \frac{\ln\ln \wt E_0/E}{4 \ln^2 \wt E_0/E}+ \ldots\r)\nn,
\ee
where the term $\sim \ln^{-2} E$ has been absorbed in the term $\sim \ln^{-1}E$ by the redefinition of the scale. This procedure fixes the scale uniquely \cite{zam95}. 

One proceeds analogously with Eqs.~\refeq{boun1}, \refeq{boun2} and obtains
\be
 T_0^{(B)}&=& \wt T_0^{(B)}\te^{\g_T^{(B)} },\,h_0^{(B)}=\wt h_0^{(B)}\te^{\g_h^{(B)}}\\
\chi_{B}(T)&=&\frac{1}{12 T \ln \wt T_0^{(B)}/T}\l(1-\frac{\ln \ln \wt T_0^{(B)}/T}{2 \ln \wt T_0^{(B)}/T}+\ldots\r)\\
\chi_{B}(h)&=&\frac{1}{4 h \ln^2\wt h_0^{(B)}/h}\l(1-\frac{\ln \ln \wt
  h_0^{(B)}/h}{\ln\wt h_0^{(B)}/h}+\ldots\r) \; .
\ee

Let us now turn to the half-filled case $n=1$. The additional Umklapp term in
\refeq{eff2} can also be bosonised and is proportional to
$\cos\sqrt{8\pi}\varphi$. When we now again rescale the field
$\varphi'=\varphi \sqrt{K_c}$ we obtain a relevant interaction
$\sim\cos\sqrt{8\pi/K_c}\varphi'$ for any finite $u>0$ because $K_c<1$ in this
case. This means that the charge sector will be massive. Indeed, at
$u\to\infty$, the excitations of the Hubbard model at half filling are exactly
those of the $XXX$-chain \cite{book}. The formulas
(\ref{suszi},\ref{boun1},\ref{boun2}) remain valid at half-filling as well.
The leading term in Eq.~\refeq{boun2}, including the constant $h_0$, was given
in \cite{AsakawaSuzuki96b}. There, a phenomenological argument was found that
generalises this result to arbitrary filling. The constant $h_0$ was left
undetermined in the arbitrary filling case.

\section{Bethe ansatz}
\label{ba}
In the framework of the Bethe ansatz (BA) solution, the energy eigenvalues of
\refeq{hdef} are given in terms of certain quantum numbers, the Bethe
roots. These roots have to be calculated from a set of coupled algebraic
equations. In the thermodynamic limit, these algebraic equations can be
transformed into linear integral equations for the densities of roots, with the energy being given by an integral over these densities. In this section, the Bethe ansatz solution is first used to verify the small-coupling expressions for $v_c,v_s,K_c$, cf. Eqs.~(\ref{vc},\ref{kc},\ref{vs}). 

Afterwards, we obtain the magnetic susceptibility at $T=0$. Therefore, we first analyse the integral equations in the small-field limit, thereby determining the constants in Eqs.~\refeq{suszi} (for $T=0$). The pre-factor $\chi_0$ has been calculated by Shiba \cite{shi72}. Our essential new results are twofold: On the one hand, the leading $h$-dependence of the bulk-susceptibility is calculated exactly at small fields, including the scale, for arbitrary fillings. On the other hand, the result for the boundary susceptibility \refeq{boun2} is confirmed within the exact solution. However, due to cumbersome calculations, the constant $\g_h^{(B)}$ in Eq.~\refeq{boun2} is left undetermined here. Finally, we consider some special cases and present numerical results. 

In order to introduce our notation, we shortly state the main results of the BA solution. For any further details, the reader is referred to \cite{book}, which also contains an extensive list of the original literature. The BA solution for the one-dimensional Hubbard model with open boundary conditions has been found by Schulz \cite{sch85}, based on the coordinate Bethe ansatz. The algebraic BA for this model has been performed in \cite{gua00}. The BA equations read
\be
\fl\te^{2\rmi k_j (L+1)} = \prod_{l=1}^{M_\down} \frac{\la_l-\sin k_j -\rmi u}{\la_l-\sin k_j +\rmi u}\frac{\la_l+\sin k_j +\rmi u}{\la_l+\sin k_j -\rmi u}\, , \;j=1,\ldots,N\label{bae3}\\
\fl\prod_{j=1}^N \frac{\la_l-\sin k_j -\rmi u}{\la_l-\sin k_j +\rmi u}\frac{\la_l+\sin k_j -\rmi u}{\la_l+\sin k_j +\rmi u} =\!\!\!\!\prod_{m=1, m\neq l}^{M_\down} \frac{\la_l-\la_m -2 \rmi u}{\la_l-\la_m +2\rmi u}\frac{\la_l+\la_m -2 \rmi u}{\la_l+\la_m +2\rmi u}\,,\nn\\
\qquad\, \; l=1,\ldots,M_\down\;,\label{bae4}
\ee
and the energy is given by
\be
E=-2\sum_{j=1}^N \cos k_j\label{gseo}.
\ee
Here, we analyse the groundstate, where the $N$-many $k_j$s and the
$M_\down$-many $\la_l$s lie on one half of the real axis, {\em except the
  origin}. Although $k=0, \la=0$ {\em are} solutions of the system
\refeq{bae3}, \refeq{bae4}, they must not be counted in \refeq{gseo}: These
solutions correspond to zero-momentum excitations, which have to be excluded
due to the broken translational invariance in the open system.\footnote{The
  wave function constructed from the coordinate BA \cite{sch85} would vanish
  identically for $k=0, \la=0$.}  However, one can show that if $k_j, \la_l$
solve \refeq{bae3}, \refeq{bae4}, then the same is true for $-k_j$,
$-\la_l$. One then ``symmetrises'' Eqs.~\refeq{bae3}, \refeq{bae4} by setting
up equations for the sets $\{p_1,\ldots,p_{2N+1}
\}:=\{-k_N,\ldots,-k_1,0,k_1,\ldots,k_{N} \} $ and
$\{v_1,\ldots,v_{2M_\down+1}
\}:=\{-\la_{M_\down},\ldots,-\la_1,0,\la_1,\ldots,\la_{M_\down} \}
$:\footnote{$k=0,\lambda=0$ are included here to introduce homogeneous
  densities of roots. Their contribution is then subtracted later on, see Eqs.~(\ref{nq},\ref{eq}).}
\be
\fl\te^{2\rmi p_j (L+1)}\frac{\sin p_j + \rmi u}{\sin p_j - \rmi u}= \prod_{l=1}^{2M_\down+1} \frac{v_l-\sin p_j -\rmi u}{v_l-\sin p_j +\rmi u}\, ,\; j=1,\ldots,2N+1\label{bae1s}\\
\fl\frac{v_l + 2 \rmi u}{v_l - 2 \rmi u}\prod_{j=1}^{2N+1} \frac{v_l-\sin p_j +\rmi u}{v_l-\sin p_j -\rmi u} =\prod_{m=1, m\neq l}^{2M_\down+1}\frac{v_l-v_m +2 \rmi u}{v_l-v_m -2\rmi u}\, ,\; l=1,\ldots,2M_\down+1\label{bae2s}.
\ee
These equations can be solved analytically in the small coupling limit,
cf. \ref{appa}. However, this solution has to be treated with care. It has
been shown in \cite{tak71} that a small-coupling expansion of the ground-state
energy {\em in the thermodynamic limit} has zero radius of convergence. This
does not come as a surprise when considering again the low-energy effective theory
presented in Sec.~\ref{ft}: At $u=0$ the interaction of vector currents
in Eq.~(\ref{hs}) changes from marginally relevant to marginally irrelevant. Thus in the following we perform the thermodynamic limit before considering any small-field or small-coupling approximations and compare with the results of \ref{appa} afterwards. 

In the thermodynamic limit, one can set up equations equivalent to \refeq{bae1s}, \refeq{bae2s}, by introducing the density of $p$s, $\rho(k)$, and the density of $\la$s, $\sigma(v)$. These densities are solutions to the following set of coupled linear integral equations \cite{sch85,AsakawaSuzuki96a}
\be
\fl\rho(k) = \frac{1}{\pi} + \frac{1}{L} \l( \frac1\pi - \cos k \,a_1(\sin k)\r) + \cos k \int_{-B}^B a_1(\sin k -v) \sigma(v) \d v \label{rho}\\
\fl\s(v)= \frac{1}{L} a_2(v) + \int_{-Q}^{Q} a_1 (v-\sin k) \rho(k) \d k - \int_{-B}^B a_2(v-w) \s(w) \d w \label{s},
\ee
where 
\be
a_n(x)=n u/(\pi(n^2 u^2 + x^2))\label{addeq}.
\ee
The integration boundaries are determined from the particle density $n$ and
the density of particles with spin down $m_\down$,
\be
n&=& \frac12 \int_{-Q}^Q \rho(k) \d k - \frac{1}{2 L} ,\; m_\down = \frac12 \int_{-B}^B \sigma(v) \d v - \frac{1}{2 L}. \label{nq}
\ee
Once these equations are solved, the energy density $e$ is obtained from 
\be
e=-\int_{-Q}^Q\cos k\, \rho(k) \d k + \frac{1}{ L}\label{eq}. 
\ee

\subsection{Velocities and Luttinger parameter}
Before proceeding further, we first make contact with the field-theoretical results (\ref{vc},\ref{kc},\ref{vs}) in the previous section. Since these concern only bulk-quantities, we discard the $1/L$-corrections in this subsection. We also set the lattice parameter $a\equiv 1$ here.  

\begin{figure}
\begin{center}
\includegraphics*[width=0.9\columnwidth]{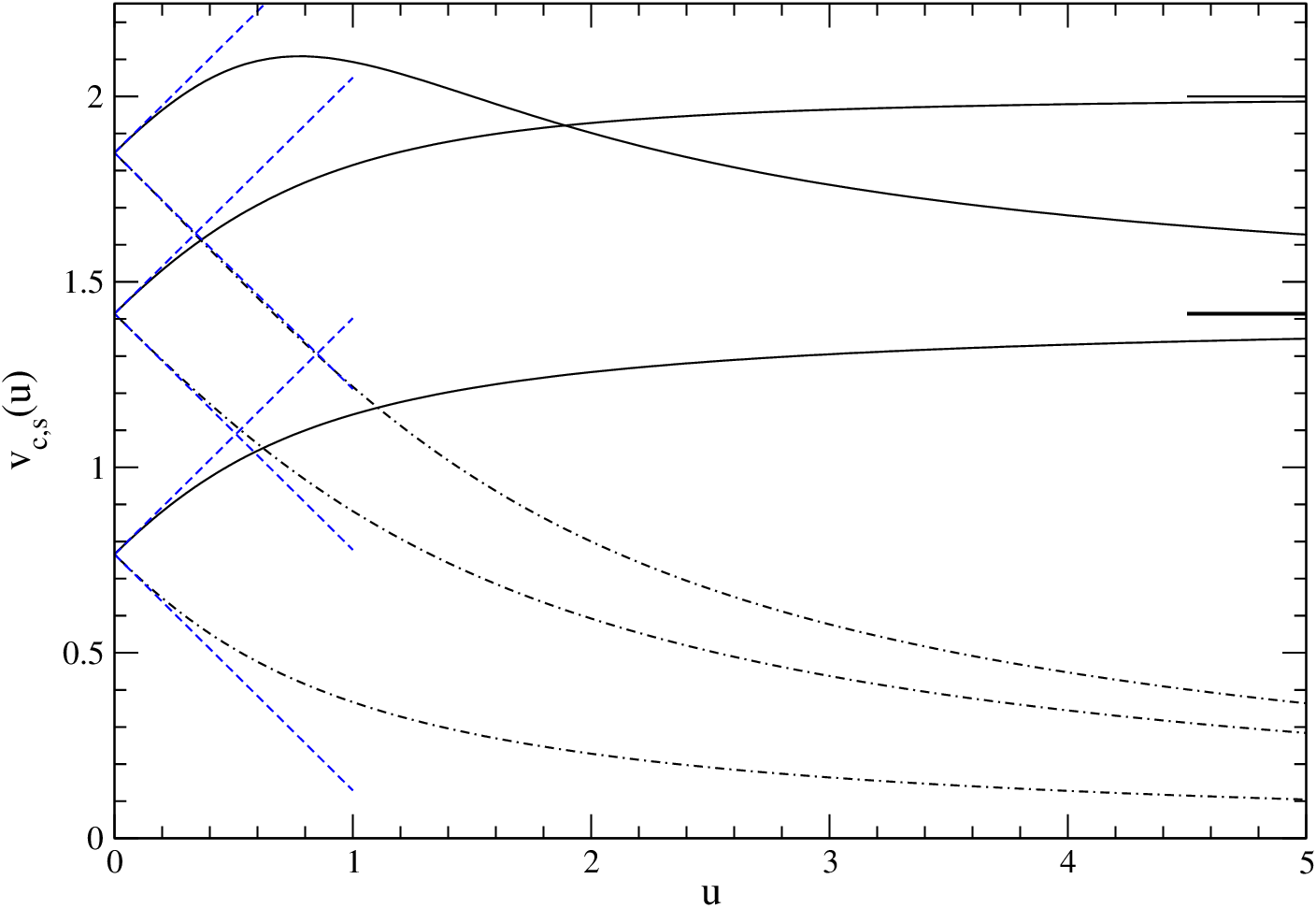}
\caption{Spin- (black point-dashed lines) and charge- (black full lines) velocities at fillings $n=1/4,\,1/2,\,3/4$ (pairs from bottom to top). The dashed lines on the left are the low-$u$ results Eqs.~\refeq{vc}, \refeq{vs}. The horizontal bars on the right indicate the limiting value $v_c|_{u\to \infty}=2 \sin (\pi n)$, \cite{sch91}. Note that this asymptotic value is the same for both $n=1/4, 3/4$. By comparing $v_c|_{u=0}=v_F=2 \sin(\pi n/2)$ with $v_c|_{u\to \infty}$, one concludes that $v_c(u)$ is maximal at a finite $u$ for $2/3<n<1$.} 
\label{figv}
\end{center}
\end{figure}

\begin{figure}
\begin{center}
\vskip0.7cm
\includegraphics*[width=0.9\columnwidth]{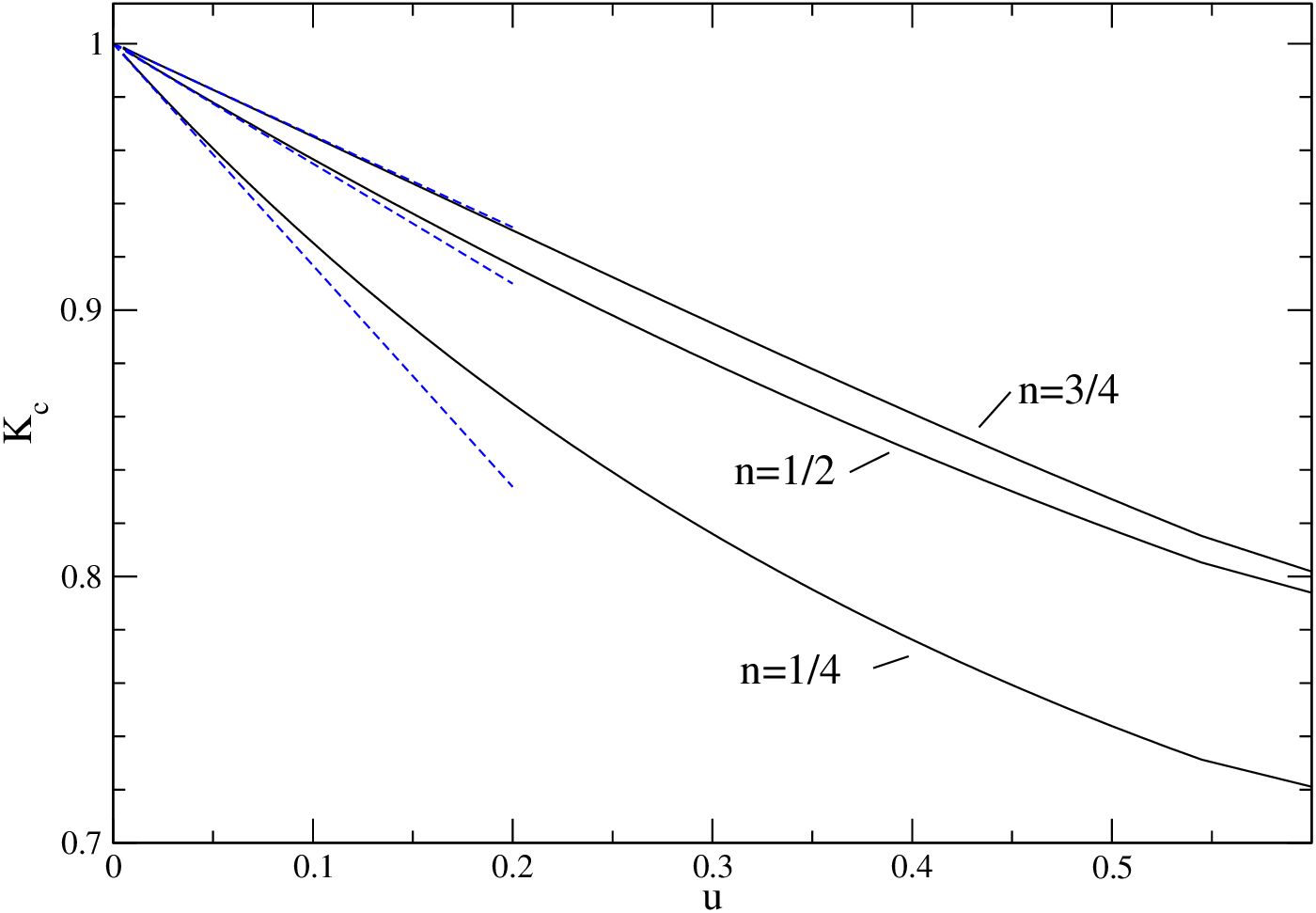}
\caption{Luttinger parameter $K_c$ at fillings $n=1/4,\,1/2,\,3/4$. The blue dashed lines are the low-$u$ results Eq.~\refeq{kc}.} 
\label{figk}
\end{center}
\end{figure}

Furthermore, the results (\ref{vc},\ref{kc},\ref{vs}) have been obtained for densities $n\neq 1$ (such that both the charge and spin channels are massless). Analogously, we restrict ourselves here to densities away from half-filling. Within the BA, charge- and spin-velocities are calculated as
\be
v_{c,s}=\frac{\partial \e_{c,s}}{\partial p_{c,s}}=\l.\frac{\partial_\la \e_{c,s}(\la)}{\partial_\la p_{c,s}(\la)}\r|_{\la=B,Q}=\l.\frac{1}{\pi}\frac{\partial_\la \e_{c,s}(\la)}{(\rho,\sigma)}\r|_{\la=B,Q}\label{vcs}
\ee
where $\e_{c,s}$ is the energy of the lowest possible (i.e. at the Fermi surface) elementary charge/spin excitation and $p_{c,s}$ the corresponding momentum. After the second equality sign, these quantities are parametrised by the spectral parameter $\la=k,v$. Then $p_{c,s}$ is expressed by the densities $\rho,\sigma$. The derivatives of the elementary excitations with respect to the spectral parameter are given by ($\e_c'(k)=\partial_k \e_c(k)$, $\e_s'(v)=\partial_v \e_s(v)$)
\be
\e'_c(k) &=& 2 \sin k + \cos k \int_{-B}^B a_1(\sin k -v) \e_s'(v) \d v \label{ec}\\
\e'_s(v)&=&  \int_{-Q}^{Q} a_1 (v-\sin k) \e'_c(k) \d k - \int_{-B}^B a_2(v-w) \e_s'(w) \d w \label{es}.
\ee
Furthermore, from $\H_c$ in Eq.~\refeq{hc} it follows that the Luttinger liquid parameter $K_c$ is obtained from the charge susceptibility $\chi_c$ at zero field,
\be
\chi_c= \frac{2 K_c}{\pi v_c}.
\ee
The susceptibility $\chi_c=\partial_\mu n$ in turn can be expressed by
$\e'(k)|_{k=B}$, $\rho(B)$ and a related function \cite{usu89}. Although the
analytical solution of the integral equations in the limit $u\to 0$ is
difficult to obtain due to singular integration kernels, these equations can
be solved numerically to high accuracy. Figs. \ref{figv}, \ref{figk} show the
velocities and the Luttinger parameter at different fillings as a function of
$u$, together with the analytical predictions (\ref{vc},\ref{kc},\ref{vs}) for
small $u$. Agreement is found in all cases.

Once $v_s$ is calculated, $\chi_0$ is also known by virtue of
Eq.~\refeq{chi0}. In the next section we describe how to obtain the
$h$-dependent corrections. 

\subsection{Spin susceptibility} 
An analytical solution of Eqs.~\refeq{rho}, \refeq{s} is a challenging task due to the finite integration boundaries $Q$, $B$. To gain a first insight, consider the integral of Eq.~\refeq{s} over the whole real axis, yielding $\int_{-\infty}^\infty \sigma(v) \d v = 2(n-m_\down) + 1/L$ and therefore the magnetisation 
\be
s:=\frac{n}{2} -m_\down = \frac{1}{2} \int_{B}^\infty \sigma(v) \d v.\label{sb}
\ee
Our aim is to perform a low-field expansion, i.e. an expansion around $s=0$. From Eq.~\refeq{sb}, this corresponds to an asymptotic expansion of $\sigma(v>B\gg 0)$. This expansion is done by generalising Shiba's approach \cite{shi72}, who calculated $\chi_0$ for pbc in a different way than via the spin velocity. 

Substitute Eq.~\refeq{rho} into Eq.~\refeq{s} to obtain
\be
\fl\sigma(v) = \frac{1}{L}(g_Q^{(0)}(v) + S_Q(v,0)) + g_Q^{(0)}(v) - \int_{-B}^B S_Q(v,v') \sigma(v') \d v' \label{sve}\\
\fl S_Q(v,v'):= a_2(v-v') - \int_{-Q}^Q a_1(v-\sin k) a_1(\sin k -v') \cos k\, \d k \nn.
\ee
Here $g_Q^{(0)}$ is the $\nu=0$-case of $g^{(\nu)}_Q$, defined by
\be
g^{(\nu)}_Q(v)&:=& \int_{-Q}^Q \frac1\pi a_1(v-\sin k) \cos ^\nu k \,\d k \label{defgn}. 
\ee
Furthermore, 
\be
\fl\sigma^{(\nu)}(v):= \frac{1}{L}(g_Q^{(\nu)}(v) + S_Q(v,0)) + g_Q^{(\nu)}(v) - \i S_Q(v,v') \sigma^{(\nu)}(v') \d v' \label{sn}.
\ee
Eq.~\refeq{sn} can be solved for $\sigma^{(\nu)}$, at the expense of introducing a new unknown function $\mathcal M_Q(v,v')$:
\be
\fl\mathcal M_Q(v,v') = S_Q(v,v') - \i S_Q(v,v'')\mathcal M_Q(v'',v') \d v''\label{ms}\\
\fl\sigma^{(\nu)}(v) = \frac{1}{L}(g_Q^{(\nu)}(v) + S_Q(v,0)) + g_Q^{(\nu)}(v)\nn\\
\fl\qquad\qquad- \i \mathcal M_Q(v,v')\l[ \frac{1}{L}(g_Q^{(\nu)}(v') + S_Q(v',0)) + g_Q^{(\nu)}(v')\r] \d v'\label{snv}. 
\ee
We consider now $\i \mathcal M_Q(v,v') \sigma(v') \d v'$, where $\sigma(v')$ is given by the rhs of Eq.~\refeq{sve}. Making use of Eq.~\refeq{snv} with $\nu=0$, one obtains
\be
\s(v)= \s^{(0)}(v) + \int_{|v'| > B} \mathcal M_Q(v,v') \s(v') \d v'\label{22} .
\ee
Similarly, starting from $\int_{-B}^B \sigma(v) \sigma^{(\nu)}(v) \d v$ (where $ \sigma^{(\nu)}(v)$ is given by Eq.~\refeq{sn}), one gets
\be
\fl\int_{|v|>B} \s(v) \s^{(\nu)}(v)\d v = \i \l[ g_Q^{(0)}(v) + \frac1L\l(g_Q^{(0)}(v)+ S_Q(v,0)\r)\r]\s^{(\nu)}(v) \d v\nn\\
\fl\qquad\qquad  -\int_{-B}^B \l\{\frac{1}{L}\l[g_Q^{(0)}(v) + S_Q(v,0)\r] + g_Q^{(\nu)}(v)\r\}\sigma(v) \d v\label{23}.
\ee
Let us now express the energy and the particle density in terms of these functions. By use of the definition \refeq{defgn}, and recalling Eqs.~\refeq{nq}, \refeq{eq}, one arrives at
\be
e-e_0 &=& -\pi \int_{-B}^B g_Q^{(2)}(v) \s(v) \d v +  \pi \i g_Q^{(2)}(v) \s^{(0)}(v) \d v\label{ee01}\\
n-n_0 &=& \frac\pi2 \int_{-B}^B g_Q^{(1)}(v) \s(v) \d v -\frac\pi2 \i g_Q^{(1)}(v) \s^{(0)}(v) \d v\label{nn01},
\ee
where $e_0$, $n_0$ are the energy and particle densities at $B=\infty$,
i.e. zero magnetic field. Note that $e,n$ are functions of both $Q,B$. First,
we hold $Q$ fixed so that both $e$ and $n$ change with varying $B$. At the end, we will account for the change in $n$ and calculate the susceptibility at fixed $n$.  

Let us now distinguish between the bulk and boundary parts in the auxiliary function $\s^{(\nu)}$ (an index $_B$ denotes the boundary contribution and should not be confused with the integration boundary $B$),  
\be
\s^{(\nu)}=:\s^{(\nu)}_{{\rm{bulk}}}+\frac1L \s^{(\nu)}_B
\ee
Then with the help of Eq.~\refeq{sn} the second term in Eq.~\refeq{ee01} is written as
\be
\fl\i  g_Q^{(2)}(v) \s^{(0)}(v) \d v= \i \l[ g_Q^{(0)}(v)\l(1+\frac1L\r) + \frac1L S_Q(v,0)\r] \s^{(2)}_{{\rm{bulk}}}(v) \d v.
\ee
The first term in Eq.~\refeq{ee01} is reformulated with the aid of Eq.~\refeq{23}. Analogous calculations are done for Eq.~\refeq{nn01}. Putting everything together allows us to write
\be
\fl e-e_0 = \pi \int_{|v|>B} \sigma(v) \sigma^{(2)}(v) \d v - \frac{\pi}{L} \int_{|v|>B} \l[ g_Q^{(2)}(v) + S_Q(v,0)\r] \s_{{\rm{bulk}}}(v) \d v\label{ee02} \\
\fl n-n_0 =- \frac\pi2 \int_{|v|>B} \sigma(v) \sigma^{(1)}(v) \d v + \frac{\pi}{2L} \int_{|v|>B} \l[ g_Q^{(1)}(v) + S_Q(v,0)\r] \s_{{\rm{bulk}}}(v) \d v\label{nn02}.
\ee
We thus have to evaluate $\sigma(v)$, $\s^{(1,2)}(v)$ asymptotically. From Eq.~\refeq{sn},
\be
\sigma^{(\nu)}(v) &=& d^{(\nu)}_Q(v)  + \i \int_{-Q}^Q \frac{a_1(\sin k -v')\,\cos k}{4 u \cosh \frac{\pi}{2 u}(v-\sin k)} \s^{(\nu)}(v') \d k\, \d v'\nn\\
d^{(\nu)}_Q(v) &:=& \l(1+\frac1L\r) \frac1\pi \int_{-Q}^Q \frac{\cos^\nu k}{4 u \cosh \frac{\pi}{2 u} (v-\sin k) }\d k \nn\\
& & + \frac1L\l( \kappa^{(1)}(v) - \int_{-Q}^Q \frac{a_1(\sin k) \cos k}{4 u \cosh \frac{\pi}{2 u} (v-\sin k)} \d k \r)\label{dqnu}.
\ee
Here the integration kernel 
\be
\kappa^{(\mu)}(p) &=& \frac{1}{2\pi} \i \frac{\te^{-\mu u|\o|}}{2 \cosh \o u} \te^{\rmi \o p} \d \o \nn
\ee
has been defined. Eq.~\refeq{dqnu} can be solved for $\sigma^{(\nu)}$, at the expense of introducing an unknown function $L_Q(t,t')$,
\be
L_Q(t,t') &=& \delta(t-t') + \int_{-\sin Q}^{\sin Q} \kappa(t-t'') L_Q(t'',t') \d t''\nn\\
\sigma^{(\nu)}(v) &=& d^{(\nu)}_Q(v)+ \int_{-\sin Q}^{\sin Q} \d t \int_{-\sin Q}^{\sin Q} \d t' \frac{L_Q(t,t')}{4\pi u \cosh \frac{\pi}{2 u }(v-t)} \nn\\
& & \times \l[ \l(1+\frac1L\r)\int_{-Q}^Q  \cos^\nu p \,\kappa^{(1)}(\sin p -t') \d p \r.\nn\\
& & \l. - \frac{\pi}{L} \int_{-Q}^Q a_1(\sin p) \cos p \,\kappa^{(1)}(\sin p - t') \d p + \frac{\pi}{L} \kappa^{(2)}(t')\r]\label{33}
\ee
From the expression \refeq{33}, the $v\to \infty$ asymptotic behaviour of $\sigma^{(\nu)}(v)$ can be read off, namely:
\be
\sigma^{(\nu)}(v)\stackrel{|v|\to \infty}{\sim} \frac1u \te^{-\frac{\pi}{2 u} |v|} I_Q^{(\nu)} + \frac1L \, \kappa^{(1)}(v)\label{34},
\ee
with an algebraic decay $\kappa^{(1)}(|v|\to \infty)\sim 1/(4 v^2)$. The quantity $I_Q^{(\nu)}$ is defined as
\be
\fl I_Q^{(\nu)}= \l(1+\frac1L\r)\int_{-Q}^Q \frac{\cos^\nu\!k}{2\pi} \,\te^{\frac{\pi}{2 u} \sin k}\, \d k - \frac{1}{2L} \int_{-Q}^Q  a_1(k)\, \cos k \,\te^{\frac{\pi}{2 u} \sin k}\,\d k \nn\\
\fl\qquad + \int_{-\sin Q}^{\sin Q} \d t\int_{-\sin Q}^{\sin Q} \frac{\d t'}{2\pi}\, \te^{\frac{\pi}{2 u} t} L_Q(t,t') \l[ \l(1+\frac1L\r)\int_{-Q}^Q \cos^\nu\!p\, \kappa^{(1)}(\sin p -t')\d p \r.\nn\\
\fl\qquad- \l.\frac1L \int_{-Q}^Q \pi a_1(\sin p)\, \cos p\, \kappa^{(1)}(\sin p - t') \d p + \frac{\pi}{L} 2\pi \kappa^{(2)}(t') \r]\label{iqn}
\ee
For later purposes, let us also separate this function into bulk and boundary parts,
\be
I_Q^{(\nu)}&=&I_{Q,{\rm{bulk}}}^{(\nu)}+\frac1L\,I_{Q,B}^{(\nu)}\nn.
\ee 
From Eq.~\refeq{22}, we now calculate $\s(v)$ in the asymptotic limit. Therefore it is helpful first to reformulate Eq.~\refeq{22} by writing 
\be
\fl\mathcal M_Q(v,v') = \kappa^{(1)}(v-v') - \int_{-Q}^Q \frac{a_1(\sin k -v') \cos k}{4 u \cosh \frac{\pi}{2 u} (v-\sin k)} \d k \nn\\
\fl\qquad+ \int_{-\sin Q}^{\sin Q} \d t  \int_{-\sin Q}^{\sin Q} \d t' \frac{1}{8 \pi u \cosh \frac{\pi}{2u}(v-t)}L_Q(t,t')\, \int_{-Q}^Q \cos^\nu \!p \kappa^{(1)}(\sin p -t') \,\d p \nn.
\ee
From this we conclude 
\be
\fl\mathcal M_Q(v+B,v'+B) + \mathcal M_Q(v+B,-v'-B) \approx \kappa^{(1)}(v-v')+ \kappa^{(1)}(v+v'+2 B).
\ee
Therefore the equation
\be
\fl\sigma(v+B) = \sigma^{(0)}(v+B) \nn\\
\fl\qquad\qquad+ \ip \l[ \mathcal M_Q(v+B,v'+B) + \mathcal M_Q(v+B,-v'-B)\r] \sigma(v'+B) \d v' \label{svb}
\ee
can be approximated by
\be
\fl \sigma(v+B) \stackrel{B\to \infty}{\sim} \frac{1}{u} \te^{-\frac{\pi}{2 u} (v+B)}I_Q^{(0)} + \frac{1}{L} \kappa^{(1)}(v+B) \nn\\
\fl\qquad\qquad\qquad+ \ip \l[ \kappa^{(1)}(v-v') + \kappa^{(1)}(v+v' + 2 B)\r] \s(v'+B) \d v'\nn\\
\fl\qquad\qquad=: \frac{1}{u} \te^{-\frac{\pi}{2 u} (v+B)}I_Q^{(0)}P_1(v+B) + \frac1L\,\frac{1}{2 u} P_2(v+B)\label{sigas}\\
\fl P_1(v)=: \te^{-\frac{\pi}{2 u} v} + \ip \l[ \kappa^{(1)}(v-v') + \kappa^{(1)}(v+v' + 2 B)\r]P_1(v') \d v'\label{p1}\\
\fl P_2(v)=: 2 u \kappa^{(1)}(v)  + \ip \l[ \kappa^{(1)}(v-v') + \kappa^{(1)}(v+v' + 2 B)\r]P_2(v') \d v'\label{p2}.
\ee
Let us estimate the error involved in the above approximations. We will see later that $B\sim -\ln h$ for small magnetic fields $h$. Corrections to the first term in Eq.~\refeq{34} are higher-order exponentials and thus would add terms $\sim h^{2 n}$ to the susceptibility (the term taken into account here yields a constant contribution $\sim h^0$). In the second term, higher-order algebraic terms have been dropped. These would contribute in order $\sim 1/(h\ln^n h)$, $n>3$, to the boundary susceptibility. The expected result \refeq{boun2} shows that all these terms are negligible for our purposes. The same holds for Eq.~\refeq{svb}. 

We continue and treat the bulk and boundary contributions separately. 

\subsection{Bulk contribution}
\label{bulkt0}
For the bulk, the quantities $I_{Q,{\rm{bulk}}}^{(\nu)}$ ($\nu=1,2$) and the function $P_1(v)$ have to be calculated. The crucial observation is that Eq.~\refeq{p1}, which determines $P_1$, is well known in the study of the spin-1/2 $XXX$-Heisenberg chain: Exactly the same function determines the $T=0$ susceptibility in that model, cf. \cite{tak99} and references therein. Eq.~\refeq{p1} is solved iteratively by the Wiener-Hopf method. The solution reads \cite{sir05b} in terms of the Fourier transform $\wt{ \mathcal{P}}(k)$ of the function $\mathcal P(v):=P(2uv)$:  
\be
\wt{ \mathcal P}(k)&=&G_+(k) \times \l\{ \begin{array}{ll} \frac{\beta_0}{k \wt B^2} , & k\neq 0 \\
 \frac{\beta_1}{\wt B}+ \beta_2 \frac{\ln \wt B}{\wt B^2} + \frac{\beta_3}{\wt B^2},& k=0\end{array} \r. \label{pk}
\ee
with $\wt B := B/(2 u)$ and
\be
\beta_0 &=& \frac{\rmi G_+(\rmi \pi) G^2_-(0)}{16 \pi^2}\,,\;\beta_1 = \frac{G_+(\rmi \pi)}{4 \pi^2} \,,\nn\\
\beta_2 &=& -\frac{G_+(\rmi \pi)}{8 \pi^3}\,,\;\beta_3 = \frac{G_+(\rmi \pi)}{8 \pi^3} \l(-\ln \pi +1\r)\nn.
\ee
The function $G_+(k)$  is given by 
\be
G_+(k)&=& \sqrt{2\pi} \frac{(-\rmi k)^{-\rmi k/(2\pi)}}{\G(1/2+\rmi
  k/(2\pi))}\,\te^{-\rmi a k}\nn\\
a&=& -\frac{1}{2\pi}-\frac{\ln(2\pi)}{2\pi}\nn.
\ee
Combining Eqs.~\refeq{ee02}, \refeq{nn02} with Eqs.~\refeq{34} and \refeq{sigas} yields the bulk magnetisation and energy in terms of the auxiliary function $\mathcal P(v)$:
\be
e-e_0 &=& \frac{4\pi}{u} \te^{-\frac{\pi}{u} B} I_{Q,{\rm{bulk}}}^{(0)} I_{Q,{\rm{bulk}}}^{(2)} \ip  \mathcal P(v) \te^{-\pi v} \d v \label{ee03}\\
n-n_0 &=& -\frac{2\pi}{u} \te^{-\frac{\pi}{u} B} I_{Q,{\rm{bulk}}}^{(0)} I_{Q,{\rm{bulk}}}^{(1)} \ip \mathcal P(v) \te^{-\pi v} \d v \label{nn03}.
\ee
Furthermore, from Eq.~\refeq{sb}, 
\be
s=\te^{-\frac{\pi B}{2 u}} I_{Q,{\rm{bulk}}}^{(0)} \ip \mathcal  P(v)\d v\label{sip}\,.
\ee
We now successively substitute $\mathcal P(v)$ from \refeq{pk} into \refeq{sip}, \refeq{ee03}, \refeq{nn03}. From the substitution of \refeq{pk} into \refeq{sip}, one obtains
\be
\wt B& =& -\frac{1}{\pi} \ln \frac{s}{s_0}\label{bs}\\
s_0 &:=&I_{Q,{\rm{bulk}}}^{(0)} G_+(0) \frac{G_+(\rmi \pi)}{\pi}.\label{s0}
\ee
Thus $e-e_0$ and $n-n_0$ are obtained as functions of $s$:
\be
e-e_0 &=& b_Q \l(\frac{s}{s_0}\r)^2 \l( 1+ \frac{b_1}{\ln s/s_0} + b_2 \frac{\ln |\ln s/s_0|}{\ln^2s/s_0} + \frac{b_3}{\ln^2 s/s_0}\r)\label{ee04}\\
n-n_0 &=& c_Q \l(\frac{s}{s_0}\r)^2 \l( 1+ \frac{b_1}{\ln s/s_0} + b_2  \frac{\ln |\ln s/s_0|}{\ln^2s/s_0} + \frac{b_3}{\ln^2 s/s_0}\r)\label{nn04}\\
b_Q &:=& \frac{4 \pi}{u} I_{Q,{\rm{bulk}}}^{(0)} I_{Q,{\rm{bulk}}}^{(2)} \frac{G^2_+(0)}{2\pi} G_+^2(\rmi \pi)\nn\\
c_Q &:=& -\frac{2 \pi}{u} I_{Q,{\rm{bulk}}}^{(0)} I_{Q,{\rm{bulk}}}^{(1)} \frac{G^2_+(0)}{2\pi} G_+^2(\rmi \pi)\nn\\
b_1&:=& 2 \beta_1 \pi/\al;\, b_2 := b_1/2\nn\\
b_3&:=& \frac{G_-(0)}{4} - \frac{2 \pi^2}{\al}\l( \beta_3 - \frac{\beta_1^2}{\al}\r) + 2 \beta_2 \pi^2\frac{\ln \pi}{\al}\nn\\
\al&:=& G_+(\rmi \pi) G_+(0)/\pi=\sqrt{\frac{2}{\pi \te}}\nn
\ee
The $Q$-dependence enters only through $b_Q,c_Q$. Since we want to calculate the susceptibility at constant density, we have to adjust $Q$ such that the density is not altered by the finite $B$-value. This adjustment is done by setting $Q=Q_0+\Delta$, where $Q_0$ corresponds to $B=\infty$:
\be
e&=& e_0 + \l( \partial_Q e_0\r) \Delta + b_Q f(s/s_0)\nn\\
n&=& n_0 + \l( \partial_Q n_0\r) \Delta + c_Q f(s/s_0)\nn,
\ee
and $f(s/s_0)$ contains the whole $s$-dependence of Eqs.~\refeq{ee04}, \refeq{nn04}. From this it follows
\be
\Delta &=& - \frac{c_Q}{\partial_Q n_0} f(s/s_0)\nn\\
e&=& e_0 +\l[ b_Q - \frac{\partial_Q e_0}{\partial_Q n_0} c_Q\r]f(s/s_0)\nn.
\ee
Now the magnetic field $h=-\partial_s e$ and the susceptibility $\chi^{-1}=\partial^2_s e$ are calculated, where $\chi$ is expressed as a function of $h$. This calculation is equivalent to considering $B$ as a variational parameter and requiring $\partial_B(e- h s)=0$, from which one obtains a relation $B=B(h)$, analogous to Eq.~\refeq{bs}. This is then substituted into Eq.~\refeq{sip} to get $s=s(h)$ and therefrom $\chi=\chi(h)$. This second approach has been chosen in \cite{sir05b}. We end up with 
\be
\chi_{{\rm{bulk}}}(h)&=& \frac{s_0}{h_0} \l( 1-\frac{1}{2 \ln h/h_0} - \frac{\ln|\ln h/h_0|}{4 \ln^2 h/h_0} + \frac{5}{16 \ln^2 h/h_0}\r) \label{chibulkt0}\\
h_0&=&\l[ b_Q - \frac{\partial_Q e_0}{\partial_Q n_0} c_Q\r]/s_0\label{h0ba}
\ee
The constant $s_0$ is given in Eq.~\refeq{s0}. Eq.~\refeq{chibulkt0} is the key result of this section for $\chi_{{\rm{bulk}}}(h)$. Note that it is of the form \refeq{suszi} (with $E=h$ there), with specified constants. Especially, an alternative expression for $v_s$ has been obtained, cf. Eq.~\refeq{chi0}: 
\be
\frac{h_0}{s_0}=2\pi v_s\label{alt}.  
\ee
To complement our discussion of the bulk susceptibility, we are going to express $h_0$ in terms of the dressed energy functions \refeq{ec}, \refeq{es} by combining Eqs.~\refeq{vcs}, \refeq{alt}. From Eqs.~\refeq{vcs},\refeq{ec},\refeq{es}, it is not difficult to show that at zero magnetic field,
\be
2\pi v_s&=& \frac{\int_{-Q}^Q \te^{\frac{\pi}{2 u} \sin k} \e_c'(k) \d k}{\int_{-Q}^Q\te^{\frac{\pi}{2 u} \sin k}\rho(k) \d k}\,.
\ee
An expression for $I_{Q,\rm{bulk}}^{(\nu)}$ equivalent to Eq.~\refeq{iqn} is as follows:
\be
I_{Q,\rm{bulk}}^{(\nu)}&=& \int_{-Q}^Q \frac{\cos^\nu\!k}{2\pi} \psi(k) \d k\label{iqpsi}\\
\psi(k)&=& \te^{\frac{\pi}{2u} \sin k} + \int_{-Q}^Q \cos k'\, \kappa^{(1)}(\sin k - \sin k')\,\psi(k') \, \d k'\label{psidef}.
\ee
Thus, by comparing with \refeq{ec}, \refeq{es} at $h=0$, 
\be
I_{Q,\rm{bulk}}^{(0)}= 2 u \l[ \sigma(v) \te^{\frac{\pi}{2 u} v}\r]_{v\to \infty}=\int_{-Q}^Q\te^{\frac{\pi}{2u} \, \sin k} \e_c'(k)\,\d k\label{iqbv}\,.
\ee
We now insert Eq.~\refeq{s0} into Eq.~\refeq{alt}, making use of Eq.~\refeq{iqbv}. Then we obtain
\be
h_0 &=& \sqrt{\frac{2}{\te \pi}} \, \int_{-Q}^Q \te^{\frac{\pi}{2 u} \sin k} \e_c'(k) \, \d k,
\ee
which is an expression equivalent to Eq.~\refeq{h0ba}.

\subsection{Boundary contribution}
\label{bount0}
Let us go back to Eqs.~\refeq{34}, \refeq{sigas}. The boundary contribution is calculated by plugging these equations into Eqs.~\refeq{ee02}, \refeq{nn02}. The resulting expression is lengthy and we do not write it down here. We rather apply the approximation to neglect terms of the order
\be
\te^{-\pi B/(2 u)}/B^2 \sim \frac{s}{\ln ^2s } \sim \frac{h}{\ln^2h}\nn
\ee
in the ground state energy. These terms would yield a contribution $\sim 1/(h\ln^3 \!h)$ to the susceptibility, cf. Eq.~\refeq{boun2}. Neglecting these terms means not fixing the scale $h_0^{(B)}$ in \refeq{boun2} uniquely. However, the algebraic $1/h$ divergence in \refeq{boun2}, dominates such terms. Then the boundary contributions read 
\be
\l[e-e_0\r]_B&=& 2\pi \ip \sigma_{{\rm{bulk}}}(v+B) \sigma^{(2)}_{{\rm{bulk}}}(v+B) \d v \label{ee0b}\\
\l[n-n_0\r]_B&=& -\pi \ip \sigma_{{\rm{bulk}}}(v+B) \sigma^{(1)}_{{\rm{bulk}}}(v+B) \d v \label{nn0b}\\
s_B&=& \frac{1}{4 u } \ip P_2(v) \d v\label{sbb},
\ee
where $P_2$ is given by Eq.~\refeq{p2}. Note that the only difference with respect to the bulk is the different dependence of $s$ on $B$, cf. Eq.~\refeq{sbb}. Instead of $s_{\rm{bulk}}\sim \exp\l[-\pi B/(2u)\r]$ we now have $s_B\sim 2u/B$, which will, according to Eq.~\refeq{bs}, cause a logarithmic divergence in the magnetisation: $1/B\sim1/\ln(h)$, yielding the divergence indicated in Eq.~\refeq{boun2}. Terms neglected in \refeq{sbb} are $\sim \exp\l[-B\r]\sim h + 1/(\ln h)$, and consequently they will yield a contribution to the boundary susceptibility which is of the bulk form, Eq.~\refeq{suszi}. These finite terms are negligible compared to the divergent terms given in \refeq{boun2}. We thus have to insert the expressions for $ \sigma_{{\rm{bulk}}}(v),\, \sigma^{(1,2)}_{{\rm{bulk}}}(v)$ from the previous section into Eqs.~\refeq{ee0b}, \refeq{nn0b}. The Fourier transform $\wt{ \mathcal P}_2(k)$ of the function $ \mathcal P_2(v):= P(2 u v)$ is given in \cite{sir05b}, namely
\be 
\wt{ \mathcal P}_2(k)&=&\l\{\begin{array}{ll}
  G_+(0)(\al_1/B+\al_2(\ln B)/B^2)& k=0\\[0.2cm]
  \rmi \al_1 G_+(k)/(k B^2),& k\neq 0
\end{array}
\r.\nn\\
\al_1&=& \frac{1}{\sqrt 2 \pi}\,,\;\al_2= -\frac{\sqrt2}{4\pi^2}\,.\nn
\ee
We now go through the same steps as in the previous subsection: Calculate $\l[e-e_0\r]_B$, $\l[n-n_0\r]_B$, $s_B$ as functions of the integration boundary $B$, and derive therefrom the boundary susceptibility as a function of $h$. This results in 
\be
\chi_B(h)&=& \frac{1}{4 h \ln^2(h_0/h)}\l(1-\frac{\ln \ln h_0/h}{\ln(h_0/h)}\r) \label{chibh}
\ee
with the scale $h_0$ given in Eq.~\refeq{h0ba}. 

\subsection{Explicit expressions in special cases}
We consider three special cases: Half filling for arbitrary coupling, as well as weak and strong coupling for arbitrary filling. 

At half filling, $Q=\pi$, $I_{Q,{\rm{bulk}}}^{(1)}=0$ and 
\be
I_{Q,{\rm{bulk}}}^{(2)}&=& \frac1\pi \int_0^\pi \sin^2k\, \te^{\frac{\pi}{2 u} \sin k} \d k \nn\\
h_0 &=& \frac{2\pi}{u} \sqrt{\frac{2\pi}{\te}}I_{Q,{\rm{bulk}}}^{(2)}.
\ee
This expression coincides with the one given by Asakawa et al. \cite{AsakawaSuzuki96b}. For strong coupling, $Q=\pi n$ and 
\be
h_0&\stackrel{u\to\infty}{\to}& \frac{n}{u} \sqrt{\frac{2 \pi^3}{\te}} \l[1-\frac{\sin(2\pi n)}{2\pi n}\r]=:h_{0,2}\label{h0lu}\\
\chi_0&\stackrel{u\to\infty}{\to}& \frac{u}{\pi^2} \l[ 1-\frac{\sin(2\pi n)}{2\pi n}\r]^{-1}\label{chilu}
\ee
The latter result has also been obtained in \cite{shi72}. It is interesting to note that
\be
h_{0,2}=\sqrt{\frac{2 \pi^3}{\te}}\, h_c\nn,
\ee
where $h_c$ is the critical field above which the system is fully polarised, $s_{\rm{bulk}}(h\geq h_c)=n/2$, \cite{book}. Thus in the strong coupling limit, the logarithmic corrections are confined to fields $h\ll h_c\sim1/u$.  In the special case $n=1$ (half filling), Eqs.~\refeq{h0lu}, \refeq{chilu} are consistent with known results for the $XXX$-chain with coupling constant $J=1/u$: For this model, $\chi_0^{(XXX)}=1/(J\pi^2)$ \cite{tak99}, and the scale $h_0^{(XXX)}=\sqrt{2\pi^3/\te}$, without taking account of the term $\sim \ln^{-2} h$ in \refeq{suszi}, \cite{sir05b}. 

The small coupling limit is technically more involved: The integration kernels in the integral equations become singular. A numerical evaluation of $\chi_0(u)$ (see section \ref{num} below) confirms the field-theoretical prediction Eq.~\refeq{chibulkhu}. As far as the scale $h_0$ is concerned, it is clear that it must diverge for $u\to 0$: Exactly at the free-fermion point $u=0$, the logarithmic corrections in Eq.~\refeq{suszi} vanish altogether; corrections to $\chi_0$ at $u=0$ are algebraic with integer powers. The leading contributions $\sim h^2,\,T^2$ for $u=0$ are calculated in the Appendix, cf. Eqs.~(\ref{ffh0},\ref{ffh0b},\ref{fft0},\ref{fft0b}). 

To describe the divergence of $h_0|_{u\to 0}$ quantitatively, consider first the half-filling case where $h_0\sim \te^{{\rm{const}}/u}$. From the general expressions of $I_{Q,{\rm{bulk}}}^{(1,2)}$, Eq.~\refeq{iqn}, it is clear that this is the case for arbitrary filling. Thus in the small-coupling limit, the finite-field susceptibility is obtained from Eq.~\refeq{suszi} with $|\ln h| \ll |\ln h_0|$, i.e. $h_0\gg h \gg 1/h_0\sim \exp[-1/u]$: 
\be
\chi_{{\rm{bulk}}}(u\to 0,h_0\gg h\gg 1/h_0)&=& \chi_0\l( 1+\frac{1}{2 \ln h_0}\r)+ \mathcal O \l(\frac{\ln h}{\ln^2\!h_0}\r)\label{chi01}.
\ee
This has to be understood such that the limit $u\to 0$ is considered at small but finite and fixed $h$. Then the $h$-dependent terms are next-leading and can be neglected in a first approximation.
On the other hand, $\chi_{{\rm{bulk}}}(u\to 0,h> 0)$ has been calculated in \ref{appa}, Eq.~\refeq{chibulkuh},
\be
\chi_{{\rm{bulk}}}(u\to 0, h>0)=\frac{1}{2\pi v_F} + \frac{2u}{\pi^2 v^2_F}\label{chi02} 
\ee
with the Fermi velocity $v_F=2 \sin (\pi n/2)$. Eqs.~\refeq{chi01}, \refeq{chi02} match provided that 
\be
h_0&=& {\rm{const }} \exp\l[ \frac{\pi}{4u} v_F\r]\label{h0su}.
\ee
We will confirm numerically this behaviour in section \ref{num}. 

\subsection{Numerical results}
\label{num}
\begin{figure}
\begin{center}
\includegraphics*[width=0.9\columnwidth]{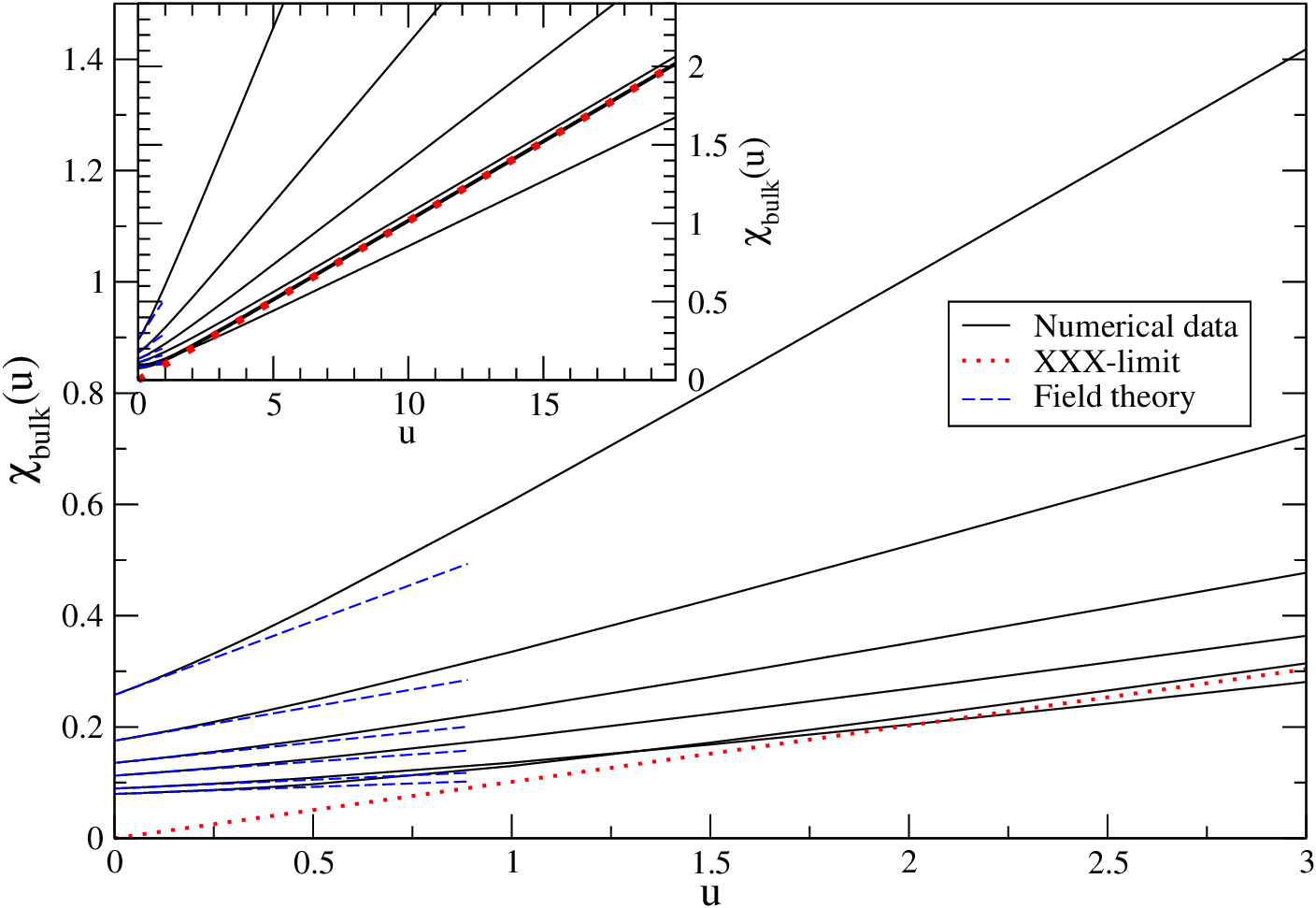}
\caption{The susceptibility $\chi_{\rm{bulk}}$ over $u$ at densities $n=0.2, 0.3,0.4,0.5,0.7,1$ (from top to bottom at $u=0$). A similar figure has been shown by Shiba \cite{shi72}. Here, we additionally compare with the field-theory result at small $u$ \refeq{chibulkhu} (blue dashed lines) and with the $XXX$-limit (for better comparison the inset shows the same figure on a larger scale; the $n=1$-line is printed bold here).} 
\label{fig1}
\end{center}
\end{figure}

\begin{figure}
\begin{center}
\includegraphics*[width=0.7\columnwidth]{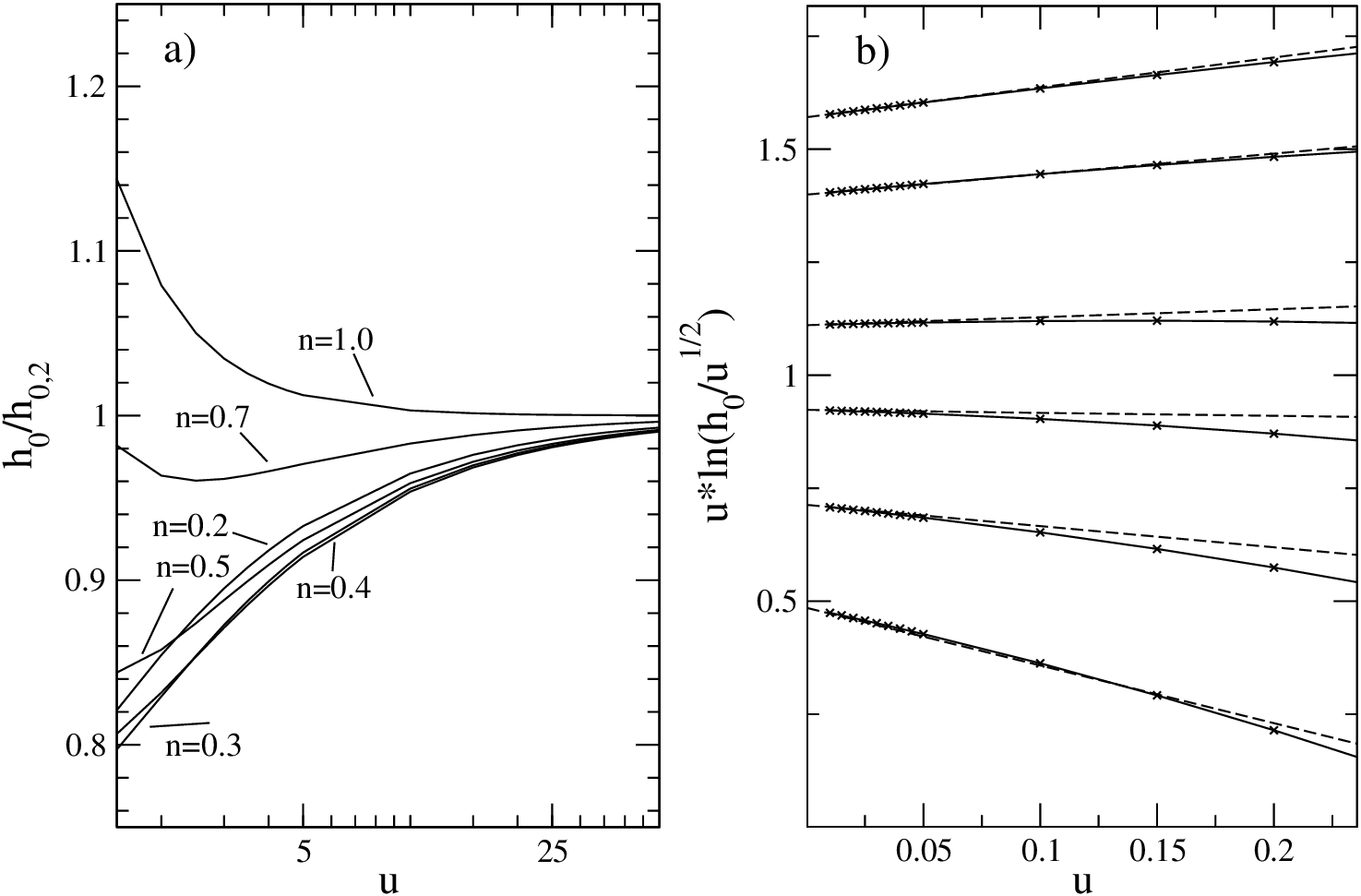}
\caption{The scale $h_0$ over $u$ for $n=0.2, 0.3, 0.4, 0.5, 0.7, 1.0$ (from bottom to top). In a), the large-coupling result Eq. \refeq{h0lu} is verified; in b), the small-coupling formula Eq. \refeq{sc} is visualised (dashed lines).} 
\label{fig2}
\end{center}
\end{figure}

In order to compare the low-field expansion Eqs.~\refeq{chibulkt0}, \refeq{chibh} to the outcome of the non-approximated integral equations \refeq{rho}, \refeq{s}, we first compute $\chi_0$, $h_0$ numerically. To do so, we follow Shiba \cite{shi72} and rewrite the quantities $\partial_Q e_0,\,\partial_Q n_0$ as solutions of linear integral equations. Namely, from Eqs.~\refeq{eq},\refeq{nq}, 
\be
\partial_Q e_0 &=& -4 \cos Q \, \rho(Q) - 2 \int_{-Q}^Q \cos k \, \partial_Q \rho(k) \d k \nn\\
\partial_Q n_0 &=& 2 \rho(Q) + \int_{-Q}^Q \partial_Q \rho(k) \d k \nn,
\ee
where $\partial_Q \rho(k)$ is obtained from Eqs.~\refeq{rho}, \refeq{s} at $h=0$ (i.e. $B=\infty$):
\be
\partial_Q \rho(k) &=& \cos k \l[ \kappa(\sin k -\sin Q) + \kappa(\sin k + \sin Q)\r] \rho(Q) \nn\\
& &+ \cos k \, \int_{-Q}^Q \kappa(\sin k - \sin p) \,\partial_Q \rho(p) \,\d p\nn.
\ee
These linear integral equations, together with Eqs.~\refeq{iqpsi}, \refeq{psidef}, are solved numerically. The result for $\chi_0$ as a function of $u$ for different fillings is given in Fig.~\ref{fig1}, together with the small-$u$ expansion \refeq{chibulkhu} and the $XXX$-limit. We checked numerically that this way of obtaining $\chi_0$ is equivalent to calculating $v_s$ from Eq.~\refeq{vcs} and then using Eq.~\refeq{chi0}. 
The scale $h_0$ as a function of $u$ at different fillings is depicted in Fig.~\ref{fig2}. Besides confirming the small-coupling result Eq.~\refeq{h0su}, we observe that the numerical data are well described by assuming the following form of the constant of proportionality in Eq.~\refeq{h0su}: 
\be
h_0 &=&\l[ \frac{4}{\pi^2} \sqrt{\frac{2\pi}{e}}\r]^{1/n} \pi \sqrt u \te^{\frac{\pi}{2 u} \sin \frac{\pi n}{2}}\label{sc}.
\ee
The exponent is exact (cf. Eq.~\refeq{h0su}), the constant is conjectured from observing good agreement with the numerics, cf. Fig.~\ref{fig2}. 
Having calculated $h_0$, $\chi_0$, the next step consists in finding $\chi(h)$ numerically and comparing with Eqs.~\refeq{chibulkt0}, \refeq{chibh}. The calculation of $\chi_{{\rm{bulk}}}(h)$ is described in \cite{book} (appendix to chapter 6). The idea is to rewrite the energy in terms of dressed energy functions (rather than in terms of dressed density functions like in Eqs.~\refeq{rho},\refeq{s}). The magnetic field enters the linear integral equations for the dressed energy functions. Once these equations are solved, both the field and the magnetisation are determined. By varying slightly the integration boundaries while keeping the density fixed, one performs a numerical derivative $\Delta s/\Delta h$ to obtain $\chi$. The results shown in the sequel demonstrate that this procedure is highly accurate. 

\begin{figure}
\begin{center}
\includegraphics*[width=0.9\columnwidth]{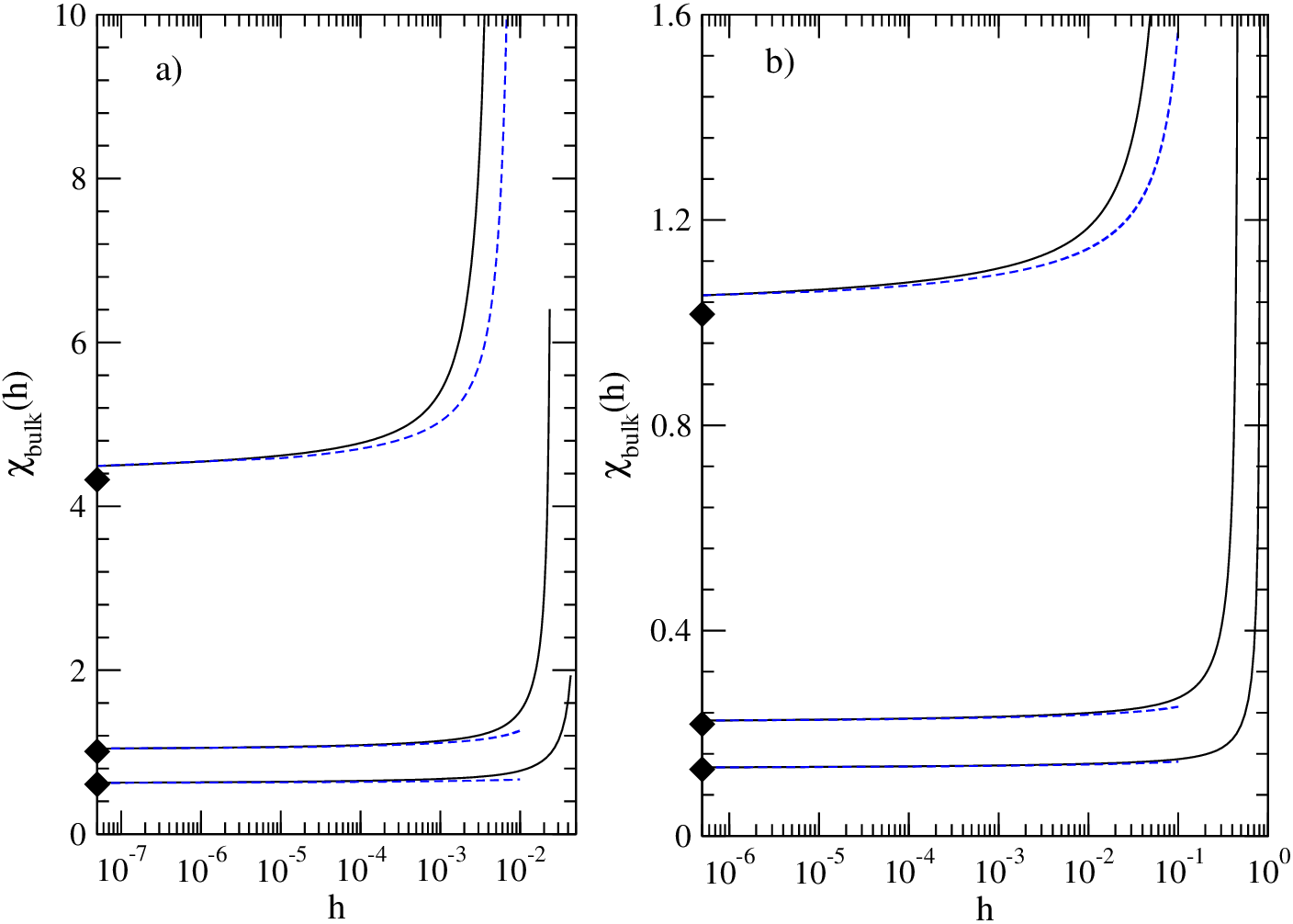}
\caption{The bulk susceptibility for $n=0.2$ in a) and for $n=1$ in b) at $u=1,2,10$ (from bottom to top). The dashed blue curves are the analytical result \refeq{chibulkt0}. The diamonds indicate $\chi_0$, showing the decrease of the scale $h_0$ with increasing $u$.} 
\label{fig3}
\end{center}
\end{figure}

\begin{figure}[b]
\begin{center}
\includegraphics*[width=0.9\columnwidth]{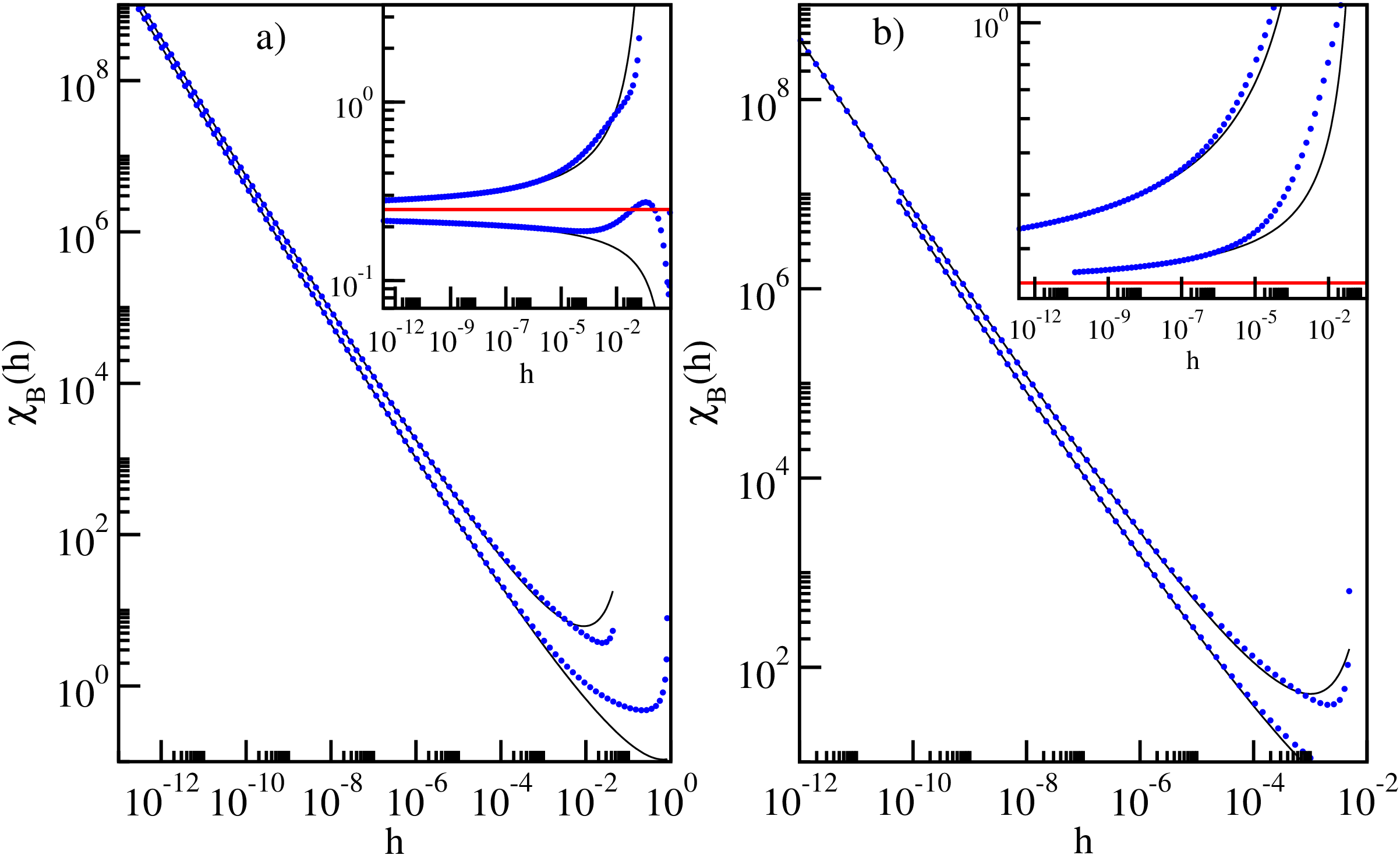}
\caption{The boundary susceptibility $\chi_B(h)$ for $u=1$ in a) and for
  $u=10$ in b) with $n=0.2,1$ (from top to bottom). The dots are the numerical
  data, the lines the asymptotic form \refeq{boun2} where $h_0^{(B)}$ has been
  determined by a fit. The insets show $\chi_B\cdot h\cdot \ln^2 h$, the red horizontal line denotes the asymptotic value $1/4$.} 
\label{fig5}
\end{center}
\end{figure}

Fig.~\ref{fig3} shows $\chi_{{\rm{bulk}}}$ for densities $n=0.2,\,1$  at different couplings, together with the analytical result \refeq{chibulkt0}.   
The boundary contribution $\chi_B(h)$ can be calculated similarly to
$\chi_{{\rm{bulk}}}(h)$ as sketched above, because the equations are linear in
the $1/L$ contribution. Results for $u=1,10$ and densities $n=0.2,\,1$
are shown in Fig.~\ref{fig5}.

\section{Friedel oscillations}
\label{corrfkts}
Due to the open boundary conditions, translational invariance is broken. This
means that one-point correlation functions like the density or the
magnetisation will no longer be just constants but rather become position
dependent. In particular, they will show characteristic oscillations near the
boundaries, the so called {\it Friedel oscillations} \cite{bed98}. These
oscillations are expected to decay algebraically with distance $x$ from the
boundary, finally reaching the mean density or magnetisation, respectively, for
$x\to\infty$.  From a field-theoretical point of view, such a one-point
correlation function can be obtained from the holomorphic part of the
corresponding two-point function \cite{Cardy84NPB}. We therefore first recall
here the asymptotics of two-point functions in the Hubbard model according to
\cite{FrahmKorepin,book}. After that, we will derive the one-point correlation
functions for the density $n(x)$ and the magnetisation $s^z(x)$. This will
then allow us to obtain the local susceptibility $\chi(x)$. By a conformal
mapping we will generalise our results to small finite temperatures. As
nuclear magnetic resonance Knight shift experiments yield direct access to
$\chi(x)$, the predictions obtained here about its position, temperature as
well as density dependence should be valuable for experiments on
one-dimensional itinerant electron systems. To test our conformal field theory
results, we present some numerical data based on the density-matrix
renormalisation group applied to transfer matrices (TMRG).
\subsection{Two-point functions}
In the following, we content ourselves with stating the results for the
asymptotics of pair-correlation functions, without giving any derivations. For
any further details, the reader is referred to \cite{book,FrahmKorepin}. We
also restrict ourselves to the case $n\neq 1$, i.e., we do not consider
half-filling. The reason to do so, is that at half-filling the charge sector
is massive and the correlation functions will be identical to those of the
Heisenberg model up to the amplitudes and the spin velocity which do depend on
$u$. The local magnetisation and susceptibility for this case, however, have
already been discussed in \cite{EggertAffleck95,BortzSirker}. 

The Hubbard model away from half-filling has two critical degrees of freedom
with different velocities and the low-energy effective theory outlined in
section \ref{ft} is therefore not Lorentz-invariant. As the spin and charge
excitations are independent from each other we can, however, still apply
conformal field theory in this situation based on a critical theory which is a
product of two Virasoro algebras both with central charge $c=1$. Due to
conformal invariance, the exponents of the correlation functions of primary
fields can then be obtained from the finite-size corrections of low-lying
excitation energies for the finite system. These in turn can be calculated
exactly {\em via} Bethe Ansatz.  Then the remaining challenge is to relate the
primary fields to the original fields of the model. This goal can be achieved
by considering the selection rules for the form factors involved and by using
additional restrictions obtained from the Bethe ansatz solution for the 
finite size spectrum \cite{FrahmKorepin}.

In this situation, the correlation function of two primary fields at zero
temperature is given by (we include here the dependence on the imaginary time
$\tau$)
\be
\fl\langle \phi_{\Delta^\pm}(\tau,x) \phi_{\Delta^\pm} (0,0)\rangle = \frac{\te^{2\rmi D_c k_{F\up} x} \te^{2 \rmi (D_c+D_s)k_{F\down} x}}{(v_c\tau + \rmi x)^{2 \Delta_c^+}(v_c\tau - \rmi x)^{2 \Delta_c^-}(v_s\tau + \rmi x)^{2 \Delta_s^+}(v_c\tau - \rmi x)^{2 \Delta_s^-}}\nn
\ee
with the scaling dimensions
\be
2 \Delta_c^\pm(\Delta\vec N, \vec D) = \l(\xi_{cc} D_c + \xi_{sc} D_s \pm \frac{\xi_{cc} \Delta N_c - \xi_{cs} \Delta N_s}{2 \det \hat \xi}\r)^2+ 2 N_c^\pm \nn\\
2 \Delta_s^\pm(\Delta\vec N, \vec D) = \l(\xi_{cs} D_c + \xi_{ss} D_s \pm \frac{\xi_{cc} \Delta N_s - \xi_{sc} \Delta N_c}{2 \det \hat \xi}\r)^2+ 2 N_s^\pm\nn. 
\ee
Let us explain the symbols used. The entries of the vector $\Delta \vec N$ are integers $\Delta N_c$, $\Delta N_s$, which denote the change of charges and down spins with respect to the ground state. The $N_{c,s}^\pm$ denote non-negative integers, and $\vec D=(D_c,D_s)$ depends on the parity of $\Delta N_{c,s}$:
\be
D_c &=& \frac12\l( \Delta N_c + \Delta N_s\r)\; {\rm{mod}} \,1\label{dc}\\
D_s &=& \frac12 \Delta N_c \; {\rm{mod}} \,1\label{ds}.
\ee
Therefore $D_c$, $D_s$ are either integers are half-odd integers. The matrix $\hat \xi$ has entries
\be
\hat \xi := \l( \begin{array}{cc} \xi_{cc}& \xi_{cs} \\ \xi_{sc} & \xi_{ss} \end{array} \r):=\l( \begin{array}{cc} Z_{cc}(B) & Z_{cs}(Q) \\ Z_{sc}(B) & Z_{ss}(Q)\end{array}\r).\label{ximat}
\ee
These entries are obtained from the following system of linear integral equations
\be
\fl Z_{cc}(k)= 1+ \int_{-B}^B Z_{cs}(v) a_1(\sin k-v) \d v\nn\\
\fl Z_{cs}(v)= \int_{-Q}^Q \cos k \, a_1(v-\sin k) Z_{cc}(k) \d k - \int_{-B}^B
   a_2(v-v') Z_{cs}(v') \d v' \nn.\\
\fl Z_{sc}(k)=\int_{-B}^B a_1(\sin k-v) Z_{ss}(v) \d v \nn\\
\fl Z_{ss}(v) = 1+ \int_{-Q}^Q \cos k \, a_1(v-\sin k) Z_{sc}(k) \d k -
   \int_{-B}^B a_2(v-v') Z_{ss}(v') \d v'\nn.
\ee
The integration kernels are given by Eq.~\refeq{addeq}.
The integration boundaries $B,Q$ are obtained from the linear integral
equations for the root densities, Eqs.~\refeq{rho},\refeq{s}.

Let us now focus onto $\langle \hat{O}(\tau,x) \hat{O}(0,0)\rangle$ with
$\hat{O}=n,s^z$, respectively. Since the operator $\hat{O}$ does neither change the particle density nor the magnetisation, we have $\Delta N_c=0 = \Delta N_s$. Consequently, according to \refeq{dc}, \refeq{ds}, $D_c=0,\pm 1,\pm2,\ldots$, $D_s=0,\pm1,\pm 2,\ldots$. Then
\be
\fl\langle \hat{O}(\tau,x) \hat{O}(0,0)\rangle - \langle\hat{O}\rangle^2 =\nn\\
\fl\qquad\;\;\;\frac{A_1\cos(2 k_{F\up} x)}{(v_c\tau + \rmi x)^{(\xi_{cc} - \xi_{sc})^2}(v_c\tau - \rmi x)^{(\xi_{cc} - \xi_{sc})^2 }(v_s\tau + \rmi x)^{(\xi_{cs} - \xi_{ss})^2}(v_s\tau - \rmi x)^{(\xi_{cs} - \xi_{ss})^2 } }\nn\\
\fl\qquad+\frac{A_2\cos \l(2 k_{F\down} x\r)}{(v_c\tau + \rmi x)^{\xi_{sc}^2} (v_c\tau - \rmi x)^{\xi_{sc}^2}(v_s\tau + \rmi x)^{\xi_{ss}^2} (v_s\tau - \rmi x)^{\xi_{ss}^2}}\nn\\
\fl\qquad+\frac{A_3\cos2(k_{F\up} + k_{F_\down})x}{(v_c\tau + \rmi x)^{\xi_{cc}^2} (v_c\tau - \rmi x)^{\xi_{cc}^2}(v_s\tau + \rmi x)^{\xi_{cs}^2} (v_s\tau - \rmi x)^{\xi_{cs}^2}}\label{sscorr}\\
\fl\qquad+\frac{A_4\cos2(k_{F\up}+2 k_{F\down})x}{(v_c\tau + \rmi
  x)^{(\xi_{cc}+\xi_{sc})^2}(v_c\tau - \rmi x)^{(\xi_{cc}+\xi_{sc})^2}(v_s\tau
  + \rmi x)^{(\xi_{cs}+\xi_{ss})^2}(v_s\tau - \rmi x)^{(\xi_{cs}+\xi_{ss})^2}
}\nn\\
\fl\qquad
+A_5\frac{x^2-v_c^2\tau^2}{(x^2+v_c^2\tau^2)^2}+A_6\frac{x^2-v_s^2\tau^2}{(x^2+v_s^2\tau^2)^2}+\cdots \nn , 
\ee 
where the amplitudes $A_i$ are different for the density-density and the
longitudinal spin-spin correlation function. The oscillating terms on the
right hand side correspond to $\vec D =(\pm 1,\mp 1),(0,\pm 1),(\pm 1,0),(\pm
1,\pm 1)$ with $\vec N = (N_c^+,N_c^-,N_s^+,N_s^-) = 0$ and the non-oscillating to $\vec N =
(1,0,0,0),(0,1,0,0),(0,0,1,0),(0,0,0,1)$ with $\vec D = 0$. 

\subsection{One-point functions in the open system}
From the two-point function $\langle
\hat{O}(z_1,\bar{z}_1)\hat{O}(z_2,\bar{z}_2)\rangle$ in Eq.~\refeq{sscorr} with $z=v_{c,s}\tau +\rmi
x$ we can read of the one-point correlation function in the presence of an
open boundary by considering it as a function of $(z_1,z_2)$ only and
identifying $z_2 = \bar{z}_1$ afterwards \cite{Cardy84NPB}.
\be
\fl \langle \hat{O}(x)\rangle - \langle\hat{O}(x)\rangle_{x\to\infty}=A_1\frac{\cos(2 k_{F\up} x+\phi_1)}{(2 x)^{(\xi_{cc} - \xi_{sc})^2}(2 x)^{(\xi_{cs} - \xi_{ss})^2} }+A_2\frac{\cos \l(2 k_{F\down} x+\phi_2\r)}{(2 x)^{\xi_{sc}^2} (2 x)^{\xi_{ss}^2}}\nn\\
\fl\qquad\qquad+A_3\frac{\cos\l[2(k_{F\up} + k_{F_\down})x+\phi_3\r]}{(2 x)^{\xi_{cc}^2} (2 x)^{\xi_{cs}^2} }+A_4\frac{\cos\l[2(k_{F\up}+2 k_{F\down})x+\phi_4\r]}{(2
  x)^{(\xi_{cc}+\xi_{sc})^2}(2 x)^{(\xi_{cs}+\xi_{ss})^2}}\label{scorr}\\
\fl\qquad\qquad+\frac{A_5+A_6}{(2x)^2} \nn ,
\ee
with unknown amplitudes $A_i$ and phases $\phi_i$ where $x$ now denotes the
distance from the boundary.

By the usual mapping of the complex plane onto a cylinder we can generalise
\refeq{scorr} to finite temperatures:
\be
\fl\langle \hat{O}(x)\rangle -
\langle\hat{O}(x)\rangle_{x\to\infty}=A_1\frac{\cos(2 k_{F\up}
  x+\phi_1)}{(\frac{v_c}{\pi T} \sinh \frac{2\pi T x}{v_c})^{(\xi_{cc} -
    \xi_{sc})^2}(\frac{v_s}{\pi T} \sinh \frac{2\pi T x}{v_s})^{(\xi_{cs} - \xi_{ss})^2} }\nn\\
\fl\;\;+A_2\frac{\cos \l(2 k_{F\down} x+\phi_2\r)}{(\frac{v_c}{\pi T} \sinh \frac{2\pi T x}{v_c})^{\xi_{sc}^2} (\frac{v_s}{\pi T} \sinh \frac{2\pi T x}{v_s})^{\xi_{ss}^2}}+A_3\frac{\cos\l[2(k_{F\up} + k_{F_\down})x+\phi_3\r]}{(\frac{v_c}{\pi T} \sinh \frac{2\pi T x}{v_c})^{\xi_{cc}^2} (\frac{v_s}{\pi T} \sinh \frac{2\pi T x}{v_s})^{\xi_{cs}^2} }\nn\\
\fl\;\;+A_4\frac{\cos\l[2(k_{F\up}+2 k_{F\down})x+\phi_4\r]}{(\frac{v_c}{\pi T} \sinh \frac{2\pi T x}{v_c})^{(\xi_{cc}+\xi_{sc})^2}(\frac{v_s}{\pi T} \sinh \frac{2\pi T x}{v_s})^{(\xi_{cs}+\xi_{ss})^2}}\label{scorrfT}\\
\fl\;\;+\frac{A_5}{(\frac{v_c}{\pi T} \sinh \frac{2\pi T x}{v_c})^2}+\frac{A_6}{(\frac{v_s}{\pi T} \sinh \frac{2\pi T x}{v_s})^2} \nn .
\ee
The magnetic susceptibility at zero field is obtained by taking the derivative with respect
to $h$ \footnote{In the same way the compressibility can be obtained by taking
  derivatives with respect to a chemical potential $\mu$.}: Here
$k_{F\up}=\pi(n+2s)/2a$, $k_{F\down}=\pi(n-2s)/2a$ and $s=\chi_0 h$ for
$|h|\ll 1$ (without
taking account of the logarithmic corrections). Furthermore, the exponents and
the amplitudes depend on $h$. However, we neglect this $h$-dependence here since it gives rise to higher-order contributions in $\chi(x)$. In the
$h=0$-case,  
\be
\xi_{cc}=:\xi,\, \xi_{ss}=1/\sqrt{2},\, \xi_{cs}=0,\, \xi_{sc} = \xi/2\nn,
\ee
leading to
\be
\chi(x)-\chi_0&=&2\pi\chi_0  x  \frac{-A_1\sin(\pi n x + \phi_1)+A_2\sin(\pi n x + \phi_2)}{(2 x)^{\xi^2/4} (2 x)^{1/2}} \nn\\
& & + 2\pi\chi_0 x \frac{A_4\sin(3\pi n x +\phi_4)}{(2 x)^{9 \xi^2/4}(2 x)^{1/2}}\label{chir1}. 
\ee
Note that the $k_{F\up}+k_{F\down}$-term in Eq.~\refeq{scorr} is
$h$-independent in lowest order and therefore does not contribute to the susceptibility. 

Again we generalise this to finite temperatures:
\be
\chi(x)-\chi_0&=& 2\pi  \chi_0 x  \frac{-A_1\sin(\pi n x + \phi_1)+A_2\sin(\pi n x + \phi_2)}{(\frac{v_c}{\pi T} \sinh \frac{2\pi T x}{v_c} )^{\xi^2/4} (\frac{v_s}{\pi T} \sinh \frac{2\pi T x}{v_s})^{1/2}}\nn\\
& & + 2\pi \chi_0 x \frac{A_4\sin(3\pi n x +\phi_4)}{(\frac{v_c}{\pi T} \sinh
  \frac{2\pi T x}{v_c} )^{9 \xi^2/4}(\frac{v_s}{\pi T} \sinh \frac{2\pi T
    x}{v_s})^{1/2}}\label{chir2}.  
\ee 
Note that we have ignored logarithmic corrections to the algebraic decay of
the correlation functions throughout this section. Multiplicative logarithmic
corrections will be present due to the marginal operator in \refeq{hs}. These
corrections have been discussed for the leading term in \refeq{sscorr} in
\cite{GiamarchiSchulz}.

\subsection{Numerical results}
\begin{figure}
\begin{center}
\includegraphics*[width=0.99\columnwidth]{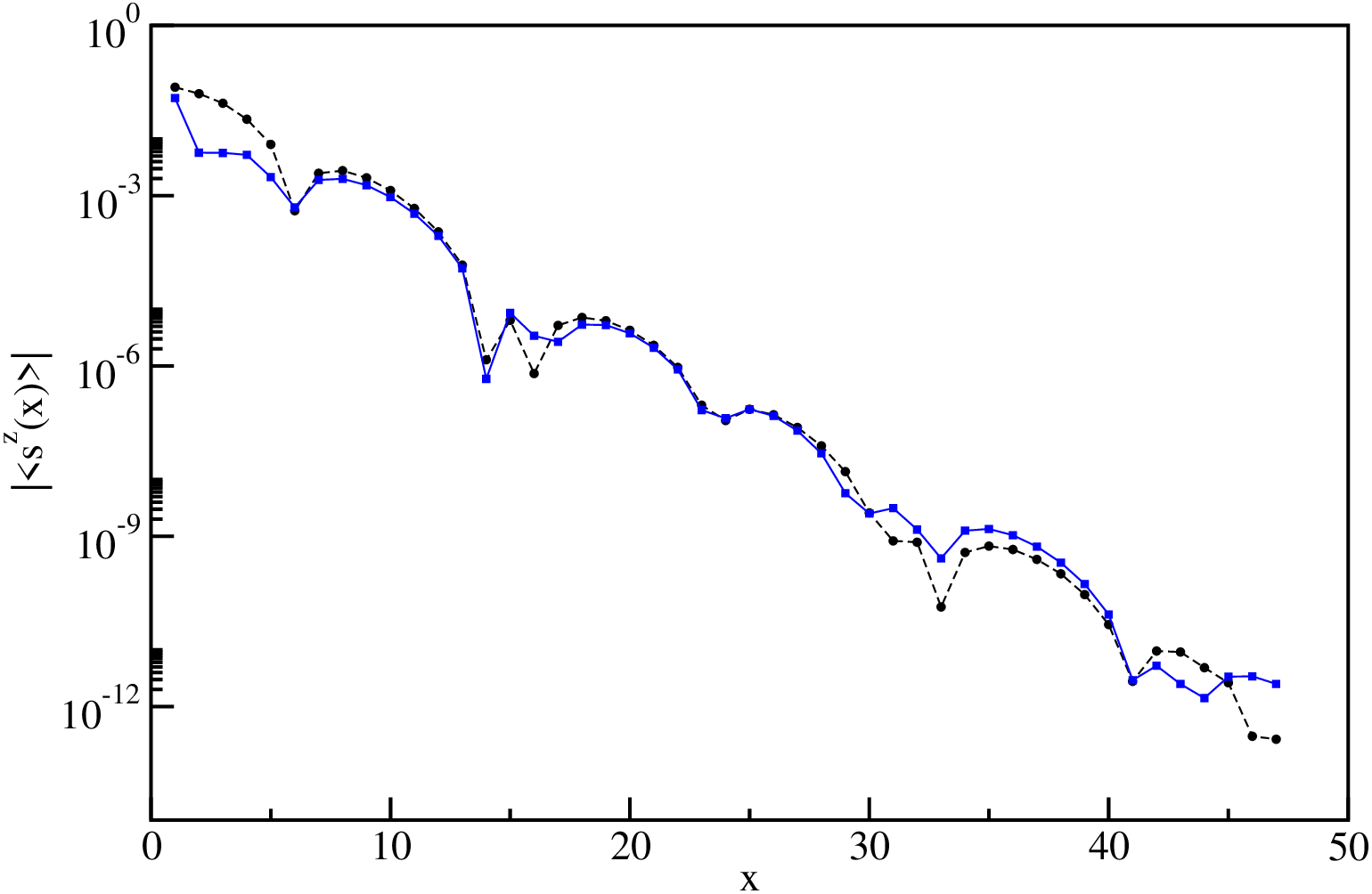}
\caption{TMRG data (black circles) for the local magnetisation $\langle s^z(x)\rangle$ where
  $u=1.0, T=0.131, s=0.037$ and $n = 0.886$. In comparison the field theory
  result (\ref{scorrfT}) is shown (blue squares) where the amplitudes and
  phases have been determined by a fit.}
\label{figmag1}
\end{center}
\end{figure}

To calculate numerically the local magnetisation $s^z(x)$ and susceptibility
$\chi(x)$ at finite temperatures we use the density-matrix renormalisation
group applied to transfer matrices (TMRG). The advantage of this method
compared to Quantum-Monte-Carlo algorithms is that the thermodynamic limit can
be performed exactly. This is particularly helpful in the present situation
where we want to study boundary effects, i.e., corrections which are of order
$1/L$ compared to bulk quantities. The method is explained in detail in
\cite{Peschel, SirkerKluemperEPL, BortzSirker}. Here we concentrate on the
local magnetisation for $h\neq 0$ and on the local susceptibility for $h=0$,
both times for generic filling $n\neq 1$. In Fig.~\ref{figmag1} TMRG data for
$\langle s^z(x)\rangle$ are shown in comparison to the field theory result
(\ref{scorrfT}). 
Here the exponents and velocities have been determined
exactly by the Bethe ansatz solutions Eqs.~(\ref{ximat},\ref{vcs}) whereas the
amplitudes and phases have been used as fitting parameters. The agreement is
very good. In particular, the exponential decay of the correlation function is
correctly described by the exponents and velocities obtained by Bethe ansatz.
The surprisingly rich structure of $\langle s^z(x)\rangle$ is caused by a
competition between the first three terms in (\ref{scorrfT}) which oscillate
with different wave vectors but have similar correlation lengths given by
$\xi_1 = 1.906, \xi_2=1.903$ and $\xi_3 = 1.708$ (the correlation lengths $\xi_i$ should not be confused with the matrix $\hat \xi$ in \refeq{ximat}).

\begin{figure}
\begin{center}
\includegraphics*[width=0.99\columnwidth]{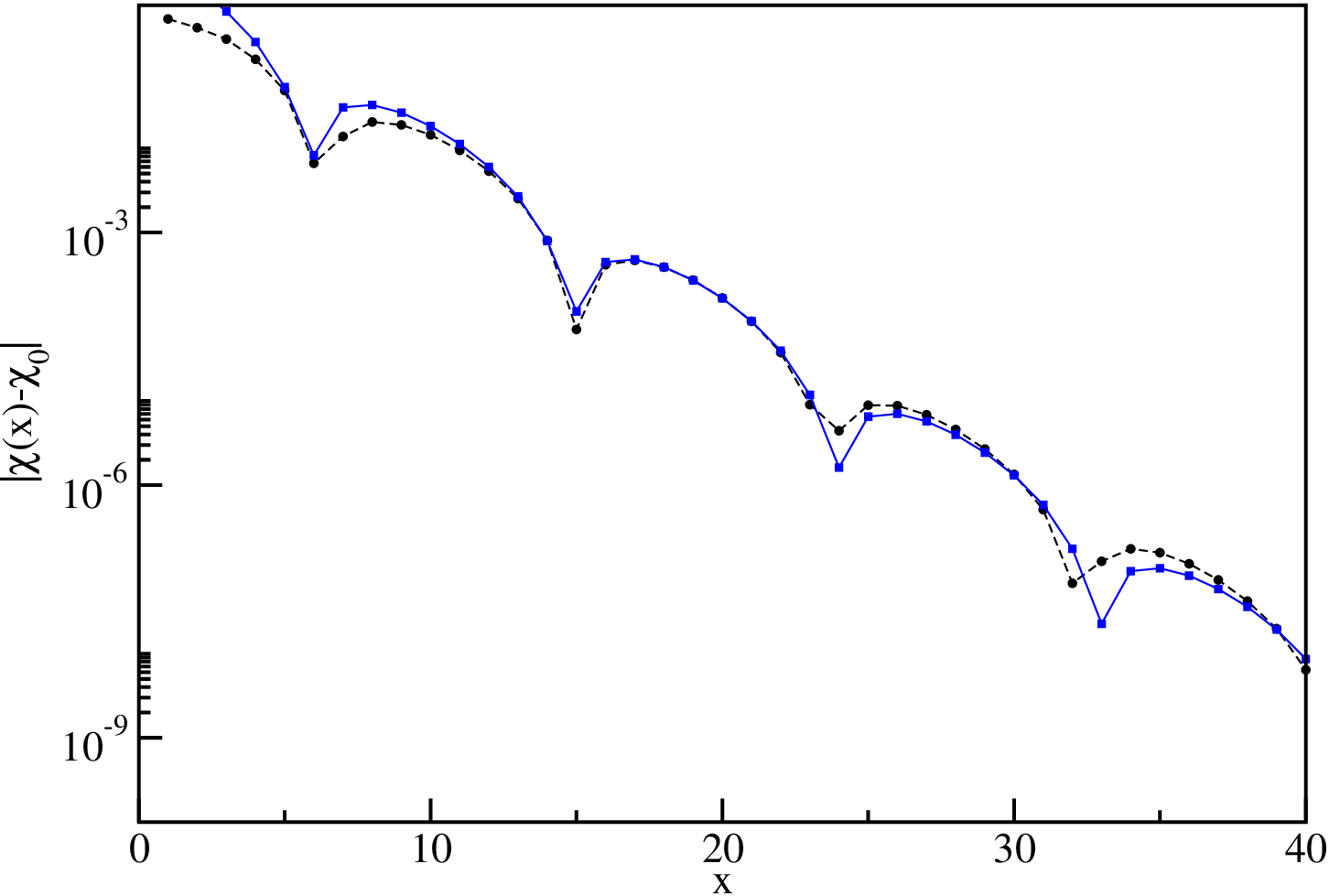}
\caption{TMRG data (black circles) for the local susceptibility $\chi(x)$
  where $u=1.0, T=0.131, s=0.0$ and $n=0.886$. In comparison the field theory
  result (\ref{chir2}) is shown (blue squares) where the amplitudes and phases
  have been determined by a fit.}
\label{fig8}
\end{center}
\end{figure}

In Fig.~\ref{fig8} the local susceptibility for the same set of parameters as
in Fig.~\ref{figmag1} but zero magnetisation is shown.
The TMRG data are again compared to the field theory result (\ref{chir2}) and
good agreement is found. Here only the first term of (\ref{chir2}) has been
taken into account because the correlation length belonging to the second term
is small compared to that of the first one.

\section{Conclusions}
We studied the low-energy thermodynamic and ground-state properties of the
one-dimensional Hubbard model with open ends. In particular, we concentrated
on the bulk and boundary parts of the magnetic susceptibility. Based on the
low-energy effective theory for this model we argued that the functional form
of $\chi_{{\rm{bulk}},B}(h,T=0)$ and $\chi_{{\rm{bulk}},B}(h=0,T)$ is universal,
i.e., does not depend on filling $n$ or interaction strength $u$. For the bulk
susceptibility only the zero temperature and zero field value $\chi_0 =(2\pi
v_s)^{-1}$ depends on $n,u$ via the spin-wave velocity $v_s$ as does the scale
$E_0$ appearing in the logarithms. For $E\ll E_0$ with $E=T,h$ , however, the
scale is not important and the bulk susceptibility becomes
$$ \chi_{\rm{bulk}} = \frac{1}{2\pi v_s}\l(1-\frac{1}{2\ln E}-\frac{\ln|\ln E|}{\ln^2 E}\r) \; .$$
For the boundary part we even find that the result for $T=0$, $h\ll h_0$ 
$$ \chi_{B} =\frac{1}{4h \ln^2 h}\l(1+\frac{\ln|\ln h|}{\ln
  h}\r) $$
as well as the result for $h=0$, $T\ll T_0$ 
$$
\chi_{B} =-\frac{1}{12T \ln T} \l(1+\frac{\ln|\ln T|}{2\ln T}\r)$$
are
completely universal. In particular, $\chi_B(T=0,h)$ and $\chi_B(T,h=0)$ are
identical to the results obtained for the Heisenberg model
\cite{FujimotoEggert,FurusakiHikihara}. The same expression for
$\chi_B(T=0,h)$ has also been obtained for the supersymmetric $t-J$ model
\cite{FHLE}. The universal behaviour of $\chi_B$ at low energies has nothing
to do with the special properties making the Hubbard model integrable.
Instead, the universal behaviour will hold for any system whose low-energy
effective theory is identical to the one for the Hubbard model described in
Sec.~\ref{ft}.  Therefore even in a generic itinerant electron system,
non-magnetic impurities or structural defects can give rise to a Curie-like
contribution to the magnetic susceptibility.  This has profound consequences
for experiments on such systems, where a Curie term in the susceptibility is
often assumed to be directly related to the concentration of magnetic
impurities in the sample. In the light of the results presented here a more
sophisticated analysis is necessary. In particular, the temperature dependence
of the Curie constant has to be analysed carefully - in the case of a boundary
considered here the Curie constant will show a logarithmic dependence on
temperature. In addition, it might be useful to investigate if the Curie
contribution can be reduced by annealing as one would expect if it is caused
by structural defects.

Based on the Bethe ansatz solution for the Hubbard model at zero temperature
we have been able to calculate $\chi_{\rm{bulk}}$ exactly beyond the limit
$h\ll h_0$ by determining the scale $h_0$ for arbitrary filling. In addition,
the exact solution has confirmed that the bulk and boundary parts show indeed
the universal functional dependence on magnetic field which has been obtained by the
low-energy effective theory.

For the Friedel oscillations in magnetisation and density caused by the open
boundaries we have derived an asymptotic expansion by making use of
conformal invariance. We have also calculated the local susceptibility near
the boundary which is a quantity directly measurable in nuclear magnetic
resonance Knight shift experiments. We confirmed our results by comparing with
numerical data obtained by the density-matrix renormalisation group applied to
transfer matrices.

\section*{Acknowledgements}
We acknowledge support by the German Research Council ({\it Deutsche
  Forschungsgemeinschaft}) and are grateful for the computing resources
provided by the Westgrid Facility (Canada). We also thank Andreas Kl\"umper
for helpful discussions.
\appendix
\section{Small-$u$ expansion}
\label{appa}
The Bethe ansatz equations for a finite system are expanded with respect to the coupling constant at small couplings, in analogy to the model of an interacting Fermi gas \cite{bat05f}. Before turning to the open boundary case, we first perform this expansion for periodic boundary conditions. The comparison with open boundary conditions yields the surface energy in this approximation.
\subsection{Periodic boundary conditions}
\label{secp}
The energy eigenvalues are given by
\be
E_{pbc}=-2\sum_{j=1}^N \cos k_j\label{gse}
\ee
where the $k_j$ are obtained through
\be
\te^{\rmi k_j L} &=& \prod_{l=1}^{M_\down} \frac{\la_l-\sin k_j -\rmi u}{\la_l-\sin k_j -\rmi u}\, ,\; j=1,\ldots,N\label{bae1}\\
\prod_{j=1}^N \frac{\la_l-\sin k_j +\rmi u}{\la_l-\sin k_j -\rmi u} &=&\prod_{m=1, m\neq l}^{M_\down} \frac{\la_l-\la_m +2 \rmi u}{\la_l-\la_m -2\rmi u}\, ,\; l=1,\ldots,M_\down\label{bae2}.
\ee
For $u=0$, $2M_\down$-many of the $k^{(0)}_j$ are grouped in pairs at $2\pi l/L$, $l=1,\ldots,M_\down$, and the $M_\down$-many $\la^{(0)}_l$s lie at $\sin (2\pi l/L )$. The rest of the $k^{(0)}_j$s (namely $N-2M_\down$ many) are not paired, they are at $\pm 2\pi j/L$, $j=-(N-M_\down-1)/2,\ldots,-(M_\down+1)/2$. We make the following ansatz, distinguishing between paired (unpaired) momenta $k^{(p)}_j$ ($k^{(u)}_j$):
\be
k^{(p)}_j&=& k^{(p,0)}_j\pm \beta_j + \delta_j^{(p)}\label{p}\\
k^{(u)}_j&=& k^{(u,0)}_j + \delta_j^{(u)}\nn\\  
\la_l&=& \la^{(0)}_l + \e_l\label{lal}.
\ee
This ansatz is motivated by evaluating numerically the BA equations \refeq{bae1}, \refeq{bae2} and it is justified a posteriori by observing that the quantities $\delta_j^{(u,p)}$, $\beta_j$, $\e_l$ can be obtained in a closed form. Eq. \refeq{p} means that two paired momenta are centred around their ``centre of mass''  $k^{(p,0)}_j + \delta_j^{(p)}$. It can be verified that $\e_l=\delta_j^{(p)}$ only for $N=2M_\down$ (no magnetisation). Expanding the BA equations \refeq{bae1}, \refeq{bae2} and comparing coefficients of the imaginary and real parts yields
\be
\fl \beta_j^2 = 2 \frac uL \cos k_j^{(p,0)}\label{beta}\\
\fl\delta_j^{(p)}= 2 \frac uL\cos k_j^{(p,0)}\l[ \sum_{l\neq j} \frac{1}{\lambda_j^{(0)} - \lambda_l^{(0)}}+\frac 12 \sum_l \frac{1}{\lambda_j^{(0)} - \sin k_l^{(u,0)}}\r] - \frac{u}{L} \frac{\sin k_j^{(0)}}{\cos k_j^{(0)}}\label{dp}\\
\fl\delta_j^{(u)}= -2 \frac uL\cos k_j^{(u,0)}\sum_l \frac{1}{\lambda_l^{(0)} - \sin k_l^{(u,0)}}\label{du}.
\ee
Note that $\beta_j^2$ may be positive or negative, depending on the value of $\cos k_j^{(0)}$. The quantity $\e_l$ in \refeq{lal} may be obtained similarly, however, it will be of no further importance for our purposes. Inserting \refeq{beta}-\refeq{du} into \refeq{gse} results in 
\be 
E_{pbc}&=& -4 \frac{\sin \frac\pi L \frac N2 \, \cos \frac \pi LS}{\sin \frac \pi2} + \frac uL (N^2-4S^2)\label{ef},
\ee
where $S=\frac{N}{2}-M_\down$. 

Now, with $n:= N/L$, $s=:S/L$,  
\be
e_{pbc}&=& -\frac4\pi \sin \frac{\pi n}{2} \cos \pi s + u\l(n^2-4s^2\r) -\frac{4\pi}{6 L^2} \sin \frac{\pi n}{2} \cos \pi s\label{gse2}.
\ee
At $s=0$ (that is, for zero magnetic field), we obtained the charge and spin velocities at small $u$ from the low-energy effective Hamiltonian in section \ref{ft}, 
\be
v_{c,s}&=& 2\sin \frac{\pi n}{2} \pm \frac{2 u}{\pi}\nn.
\ee
This provides a consistency check on the leading finite-size correction of the ground state energy, as obtained from conformal field theory \cite{BloeteCardy}  
\be
e_{pbc}:= \frac{E_{pbc}}{L} = e^{(\infty)} -\frac{\pi}{6 L^2} (v_c+v_s)  \label{eg}.
\ee
At finite magnetic field 
\be
h=-\partial_s e_{pbc}=-4 \sin\frac{\pi n}{2} \sin\pi s+2 u s\label{magf},
\ee
the susceptibility is derived from Eq.~\refeq{gse2},
\be
\chi_{{\rm{bulk}}}(u\to 0,h)=\frac{1}{4 \pi \sin \frac{\pi n}{2} \,\cos \pi s} \l( 1+ \frac{2 u }{\pi \sin \frac{\pi n}{2} \,\cos \pi s}\r).\label{chibulklin}
\ee
It is important to note that Eq.~\refeq{chibulklin} has been derived at finite $s$, in the limit of vanishing $u$. In the limit of small fields, $s\propto h$, and thus the small-field expansion of \refeq{chibulklin} reads
\be
\chi_{{\rm{bulk}}}(u\to 0,h> 0)=\frac{1}{2\pi v_F} + \frac{2u}{\pi^2 v^2_F}\label{chibulkuh},
\ee
with $s=s(h)$. The order of the limits is important here: The small-coupling limit has been taken {\em before} the small $h$-limit. That is, Eq.~\refeq{chibulkuh} is valid at small but still finite fields, where the field-dependent terms have been neglected (in the main part, cf. Eq.~\refeq{chi01}, a lower bound on the field is given in terms of the scale $h_0$, namely $h\gg 1/h_0$). These singular limits are due to the non-analytic behaviour of $\chi_{\rm{bulk}}$ as a function of the magnetic field in the thermodynamic limit, cf. Eq.~\refeq{boun2}. 

The result for $\chi_{{\rm{bulk}}}(h\to 0, u> 0)$, that is, with interchanged limits compared to Eq.~\refeq{chibulkuh}, is obtained from the low-energy effective Hamiltonian given in section \ref{ft}:
\be
\chi_{{\rm{bulk}}}(h\to 0, u> 0)\equiv \chi_0&=&\frac{1}{2\pi v_s}\nn\\
&=&\frac{1}{2\pi v_F} + \frac{u}{\pi^2 v^2_F} \label{chibulkhu}
\ee
where the result for $v_s$ given in Eq.~\refeq{vs} has been inserted. The origin of the difference between \refeq{chibulkuh} and \refeq{chibulkhu} is clarified in section \ref{bulkt0}. 

\subsection{Open boundary conditions}
For open boundary conditions, the BA equations are given in \refeq{bae3}, \refeq{bae4}. The remarks from \ref{secp} apply similarly to this case, with the modification that all roots are on one half of the real axis. Furthermore, the above expansion procedure can be repeated with the results 
\be
\beta_j^2 &=&  \frac{u}{L+1} \cos k_j^{(p,0)}\label{betao}\\
\delta_j^{(p)}&=& \frac{u}{L+1}\cos k_j^{(p,0)}\l[ \sum_{l\neq j} \frac{1}{\lambda_j^{(0)} - \lambda_l^{(0)}}+\frac 12 \sum_l \frac{1}{\lambda_j^{(0)} - \sin k_l^{(u,0)}}\r] \nn\\
& &- \frac{u}{2(L+1)} \frac{\sin k_j^{(0)}}{\cos k_j^{(0)}}\label{dpo}\\
\delta_j^{(u)}&=& - \frac{u}{L+1}\cos k_j^{(u,0)}\sum_l \frac{1}{\lambda_l^{(0)} - \sin k_l^{(u,0)}}\label{duo}.
\ee
Here the sums run over the symmetrised sets of BA-numbers. We now obtain the energy
\be
\fl E_{obc}=-2 \frac{\sin\frac{\pi(N+1)}{2(L+1)} \cos\frac{\pi S}{L+1}}{\sin \frac{\pi}{2(L+1)}}+2+\frac{u}{L+1}(N^2-4S^2 +N-2S)\label{eob}.
\ee 
Expanding \refeq{eob} in powers of $1/L$ yields the boundary contribution to the energy in this weak-coupling approximation 
\be
e_B&=&\sin\frac{n \pi}{2} \l(-4 s \sin s\pi -\frac 4\pi \cos s\pi\r) + 2(n-1) \cos s\pi \cos\frac{n \pi}{2} +2\nn\\
& &+u(2s-n)(2s+n-1)\label{eblin}.
\ee
If the boundary susceptibility is derived from this expression as in the previous section, one would obtain a constant depending on $n$ and $u$ only. This result cannot be related to Eq.~\refeq{boun2}, showing again the non-commutativity of diverse limits at $u\neq 0$.

\section{Free Fermions}
In this section we give $\chi_{\rm{bulk}}$, $\chi_B$ for free fermions ($u=0$) in the low-energy limit. The corresponding quantities are marked by an index $^{(ff)}$. 

At $T=0$, $\chi^{(ff)}(h)$ is directly obtained from Eqs.~\refeq{gse2} (the bulk part) and \refeq{eblin}, both at $u=0$, with the magnetic field given by $h=-\partial_s e$. In the small-field limit, one obtains
\be
\fl\chi_{\rm{bulk}}^{(ff)}(h)&=&\frac{1}{2\pi v_F} + \frac{1}{16\pi v_F^3}h^2+ \mathcal O(h^4)\label{ffh0}\\
\fl\chi_B^{(ff)}(h)&=&\frac{1}{2\pi v_F} + \frac{(n-1) \cos\frac{n \pi}{2}}{2 v_F^2}+ \frac{h^2}{16\pi} \l( \frac{1}{v_F^3} + \frac{\pi (n-1) \cos\frac{\pi n}{2}}{v_F^4}\r)+ \mathcal O(h^4)\label{ffh0b}.
\ee
Note that $\chi_{\rm{bulk}}^{(ff)}(h)=\chi_B^{(ff)}(h)$ only for $n=1$. 

To calculate the susceptibility at finite temperatures, one starts with the free energy per lattice site $f^{(ff)}$, 
\be
\fl-\beta f^{(ff)} = \frac{1}{L} \sum_{j=1}^L \l[\ln \l( 1+ \exp\l[-\beta\l(-2 \cos \frac{\pi j}{L+1} -\mu - h/2\r)\r]\r) + (h\leftrightarrow -h)\r] \label{intff},
\ee
where the chemical potential $\mu$ is to be determined from $n=-\partial_\mu f^{(ff)}$. Applying the Euler-MacLaurin formula to Eq.~\refeq{intff} yields
\be
\fl -\beta f^{(ff)} =\l(1+\frac1L\r)\frac1\pi \int_{0}^\pi \ln\l( 1+ \te^{-\beta(-2 \cos k -\mu - h/2)}\r) \d k \nn\\
\fl\qquad\qquad-\frac{1}{2L}\l[\ln\l(1+\te^{-\beta(-2-\mu-h/2)}\r)\l(1+\te^{-\beta(2-\mu-h/2)}\r)\r] + (h\leftrightarrow -h)\label{intfff}.
\ee
In the $T=0$-limit, the results Eq.~\refeq{ffh0}, \refeq{ffh0b} are recovered. By performing a saddle-point approximation around the two Fermi points in the integral in \refeq{intfff}, one obtains the first $T$-dependent correction to the zero-field susceptibility,
\be
\fl\chi_{\rm{bulk}}^{(ff)}(T)= \frac{1}{2\pi v_F} +\l[\frac{2\pi}{3 v_F^5}\l(1-\frac{1}{4} v_F^2\r)+\frac{\pi}{12 v_F^3}\r] T^2+ \mathcal O(T^4)\label{fft0}\\
\fl\chi_{B}^{(ff)}(T)= \frac{1}{2\pi v_F} + \frac{(n-1) \cos\frac{n \pi}{2}}{2 v_F^2}+\l[\frac{2\pi}{3 v_F^5}\l(1-\frac{1}{4} v_F^2\r)+\frac{\pi}{12 v_F^3}\r.\nn\\
\fl\qquad\qquad+\l.\frac{\pi^2}{3}(n-1)(7+3\cos n\pi)\frac{\cos n\pi}{v_F^6}+\frac{\pi^2 (n-1) \cos\frac{n\pi}{2}}{3 v_F^4}\r] T^2+ \mathcal O(T^4)\label{fft0b}.
\ee
As in the $T=0$-case, $\chi_{\rm{bulk}}^{(ff)}=\chi_B^{(ff)}$ for $n=1$ only. 

\section*{References}

\end{document}